# Interdisciplinary PhDs face barriers to top university placement within their disciplines

Xiang Zheng[1], Anli Peng[2], Xi Hong[1], Cassidy R. Sugimoto[3], Chaoqun Ni[1*]

[1] Information School, University of Wisconsin—Madison, Madison, Wisconsin, 53706, USA

[2] ADVANCE Program, University of Michigan, Ann Arbor, Michigan, 48105, USA

[3] School of Public Policy, Georgia Institute of Technology, Atlanta, Georgia, 30332, USA

* Chaoqun Ni. Email: chaoqun.ni@wisc.edu

**Abstract**

Interdisciplinary research has gained prominence as a necessity for addressing complex challenges, yet its impact on early academic careers remains unclear. This study examines how interdisciplinarity during doctoral training influences faculty placement at top universities across diverse fields. Analyzing the career trajectories of over 30,000 tenure-track faculty members who earned their Ph.D. degrees after 2005 and their initial faculty placement at 355 U.S. universities, we find that faculty newly hired by top-ranked universities tend to be less interdisciplinary in their Ph.D. research, particularly when they obtained Ph.D. from top universities and remain in their Ph.D. research field. This may reflect community trends towards homogeneity: at top universities, the existing faculty research is less interdisciplinary and more aligned with the candidates that they hire (who also exhibit lower interdisciplinarity). This preference disadvantages the placement of women graduates, who exhibit higher interdisciplinarity on average. Furthermore, we show that newly hired faculty with greater interdisciplinarity, when placed at top universities, tend to achieve higher long-term research productivity. This suggests a potential loss in knowledge production if top universities continue to undervalue interdisciplinary candidates. These findings highlight structural barriers in faculty hiring and raise concerns about the long-term consequences of prioritizing disciplinary specialization over interdisciplinary expertise.

## Introduction

Interdisciplinarity has become increasingly essential for addressing complex, global challenges that exceed the capacity of a single field[1–3]. In the modern academic system, as scholars engage in new topics and methodologies to establish unique research niches, disciplinary boundaries are constantly being contested and reconfigured[4]. Researchers with interdisciplinary expertise leverage diverse backgrounds and bring innovative methodologies and perspectives to their work that isolated disciplinary silos may miss[5]. Evidence has shown the significant impact of interdisciplinary research on technological advancements and sustained scientific impact over time[6,7]. Recognizing its significance, funding agencies like the National Institute of Health, the National Science Foundation, and the European Commission actively prioritize interdisciplinary projects to address societal challenges and enhance national innovation[8–11].

To train interdisciplinary researchers for the scientific workforce, doctoral education plays a crucial role[12]. In response to the increasing demand for interdisciplinary scholars, many universities have taken action to promote interdisciplinary research in U.S. colleges and universities[13,14], including recruiting faculty with high interdisciplinary backgrounds and building interdisciplinary research clusters[15,16]. Universities have also proactively reformed their doctoral programs or established new interdisciplinary programs to facilitate curriculum and research training to enhance the Ph.D. students' problem-solving skills toward complex, multifaceted challenges[17,18].

Interdisciplinarity presents both opportunities and challenges for the career development of early-career researchers. On the one hand, interdisciplinarity integrates diverse expertise, enhancing adaptability, problem-solving skills, and other transferable skills that are critical for academic careers[19–21]. Engaging in interdisciplinary research can also yield direct benefits, such as a wider readership across diverse scholarly communities[22,23] and increased impact[8,9,24]. Over the long term, interdisciplinary researchers may surpass their specialized counterparts in funding acquisition[25]. However, early-career researchers with high interdisciplinarity may face significant hurdles in research performance, including lower productivity, reduced funding success, and lower salaries[25–27]. The citation impact of interdisciplinary research is often delayed and difficult to translate into immediate benefits[28]. Interdisciplinary researchers also face career challenges such as limited institutional support, extra time and effort requirements due to the cognitive complexity, and difficulty in framing their work[18,24]. The prevailing disciplinary structures, including academic communities and journals, often favor specialized studies and lack the breadth of knowledge necessary to properly evaluate interdisciplinary research[29,30]. Consequently, interdisciplinary scholars often struggle with establishing a clear academic identity, which can hinder their professional visibility and integration within academic communities[15,31]. Therefore, it is crucial to understand how and why interdisciplinarity influences the career prospects of early-career researchers.

This study focuses on one aspect of interdisciplinarity's influence on scholars' careers. Specifically, we explore the relationship between initial interdisciplinarity derived from one's doctoral education training (hereafter, Ph.D. interdisciplinarity) and early academic career outcomes, particularly the attainment of tenure-track faculty jobs at universities. We operationalize Ph.D. interdisciplinarity using the Rao-Stirling index[2], a widely accepted measure that captures the disciplinary diversity of knowledge sources in a scholar's research by analyzing the referenced disciplines in their Ph.D.-stage publications. This index reflects the breadth and integration of disciplinary knowledge during doctoral training (see **Supplementary Fig. S1** for the distribution). Concerns have been raised about structural barriers in faculty hiring for interdisciplinary scholars, as departmental structures evolve slowly, often leading universities to prioritize candidates with well-defined disciplinary alignment[13,18,32]. In contrast, given the recognized benefits of interdisciplinarity and universities' sustained efforts to promote it, it is also plausible that interdisciplinarity enhances Ph.D. graduates' chances of securing faculty positions at prestigious universities[33]. However, the extent to which this trend persists, its magnitude, and its broader implications for doctoral education, faculty hiring, and knowledge production remain unclear.

To answer these questions, we examined the Ph.D. training backgrounds and career trajectories of 31,775 faculty members and their initial faculty placement at 355 U.S. Ph.D.-granting universities. Specifically, our dataset includes individuals who earned their Ph.D. degrees between 2005 and 2018 and later secured their initial tenure-track faculty positions at a U.S. university. Our analysis encompasses five broad fields: Math & Computing, Physics & Engineering, Life & Earth, Biology & Health, and Social Sciences, as well as 24 subfields categorized based on the department information of faculty appointments (see **Supplementary section 1**). Sample distribution and descriptive statistics are provided in **Supplementary Table S1.**

## Results

### Lower Ph.D. interdisciplinarity among faculty at top universities

To understand how Ph.D. interdisciplinarity relates to faculty placement, we begin by examining general patterns in faculty hiring and doctoral training. Following previous research[34], we empirically calculated the "faculty production rank" of each US university within each subfield using the AARC data, demonstrating the capability to place a university's Ph.D. graduates as faculty in other universities. Each U.S. university is assigned rank percentiles computed within subfields, where a higher percentile indicates a higher ranking. Consistent with previous findings of faculty hiring[34], most faculty's initial placement ranked below their Ph.D.-granting universities (see **Supplementary Fig. S1**). Specifically, in our sample, 59.1% of new hires were placed at universities ranked more than 10 percentiles lower than their Ph.D. institutions, while only 6.5% secured positions at universities ranked more than 10 percentiles higher. Furthermore, the proportion of new faculty hired by top universities has remained stable or slightly declined over the period, suggesting the persistent challenge of obtaining faculty positions at top universities (see **Supplementary Fig. S2**).

Examining rank and interdisciplinarity reveals that faculty newly hired by top universities demonstrate significantly lower levels of Ph.D. interdisciplinarity compared to those hired at non-top universities during the same period. As shown in **Fig. 1a**, the average Ph.D. interdisciplinarity remains stable at around 0.45 for faculty employed at universities ranking between the 0th and 80th percentiles across four fields. However, starting from the top 20% of universities (ranked above the 80th percentile), the average Ph.D. interdisciplinarity declines as the university rank increases, suggesting that top universities prefer hiring new faculty with lower levels of Ph.D. interdisciplinarity. For the most prestigious, top 5% universities, their new hires' average Ph.D. interdisciplinarity is lower than that at lower-ranked universities by 0.03 in Physics & Engineering (Welch *t*-test $p<0.001$), 0.02 in Life & Earth Sciences ($p=0.028$), 0.03 in Biology & Health ($p<0.001$), and 0.03 in Social Sciences ($p<0.001$). Math & Computing stands out as an exception ($p=0.784$). In practical terms, this means that top universities are more likely to select candidates whose Ph.D. research stays closer to established disciplinary boundaries rather than those who integrate knowledge across fields.

Additionally, the proportion of faculty at top universities typically decreases as their Ph.D. interdisciplinarity increases (see **Supplementary Fig. S3**). Comparing the yearly average paper-level interdisciplinarity of faculty at top universities and other universities during their Ph.D. stage, we found that although both groups exhibited increasing interdisciplinarity over time, faculty who were placed at top universities had lower levels of interdisciplinarity in the Ph.D. stage for most of the years (see **Supplementary Fig. S4**). This suggests potential resistance among top universities to hiring scholars with broader interdisciplinary backgrounds.

**Fig. 1. Newly hired faculty's Ph.D. interdisciplinarity is lower at top universities.** Ph.D. interdisciplinarity is defined here as an individual's median Rao–Stirling index over their Ph.D.-period papers. Results based on a Rao-Stirling index derived from the pooled references of those papers are shown in **Supplementary Fig. S16**. **a.** Average Ph.D. interdisciplinarity across faculty universities rank percentiles by fields. Trend lines were generated using LOWESS smoothing, with shaded areas representing 95% bootstrap confidence intervals. **b.** Odds ratios of standardized Ph.D. interdisciplinarity for faculty placement at top 10% universities. We estimated a series of Logistic regression specifications with various combinations of controls (Set 1: Ph.D. university rank and Ph.D. graduation year; Set 2: Set 1+personal and performance features; Set 3: Set 2+advisor features; see **Data and methods**). *** $p < 0.001$, ** $p < 0.01$, * $p < 0.05$. See **Fig. S5a** for regression results using other cutoffs (5%, 15%, and 20%) of top university rank percentiles.

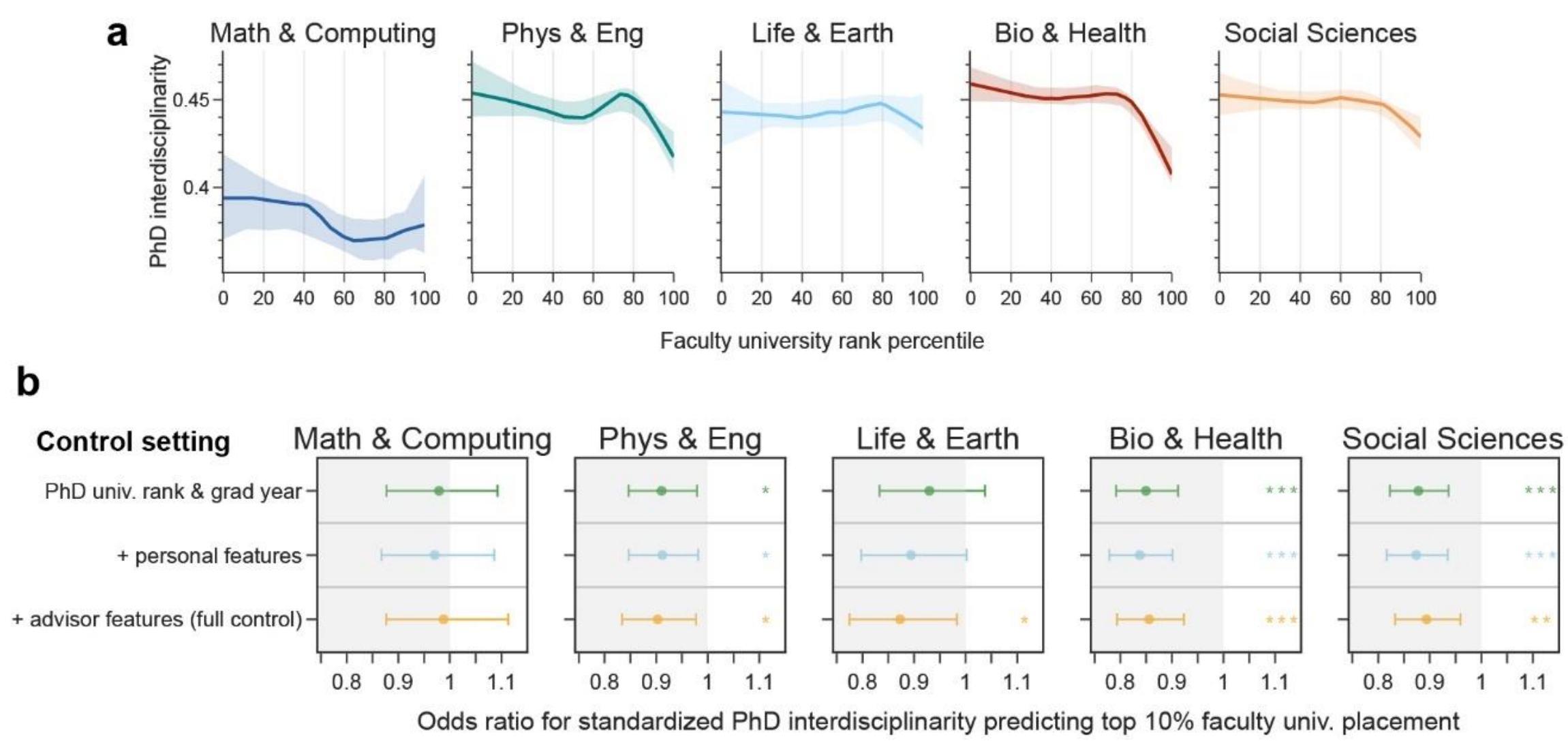

To further explore this relationship, we conducted a series of logistic regressions to examine the relationship between Ph.D. interdisciplinary and the likelihood of securing a faculty position at top universities. We controlled for different sets of variables to account for potential confounding factors. Specifically, each model accounted for the subfield-specific rank percentile of the university where an individual received their Ph.D. degree and the Ph.D. degree-awarded year. Additional models adjusted for variables related to personal features, performance, collaborators, and Ph.D. advisors' features (see **Data and methods**). As shown in **Fig. 1b**, except for Math & Computing, higher levels of Ph.D. interdisciplinarity are associated with a lower likelihood of securing faculty positions at universities ranked in the top 10%. Specifically, under the full controls, a 1 standard deviation (SD) increase in Ph.D. interdisciplinarity reduces 10% in the odds of securing a faculty position at a top 10% university in Physics & Engineering (OR=0.90, 95% CI [0.83, 0.97], $p$=0.011), 13% in Life & Earth Sciences (OR=0.87, 95% CI [0.77, 0.98], $p$=0.025), 15% in Biology & Health (OR=0.85, 95% CI [0.79, 0.92], $p$<0.001), and 11% in Social Sciences (OR=0.89, 95% CI [0.83, 0.95], $p$=0.002). We also varied the cutoff rank percentile for top universities to 5%-20% and found similar results in 5% and 15% (see **Supplementary Fig. S5a**). Using the raw, untransformed Ph.D. interdisciplinarity measure produces similar results (see **Supplementary Fig. S6a**). Notably, the negative association diminishes as the definition of top universities expands to include a larger pool of universities.

As a robustness check, we examined how the relationship between Ph.D. interdisciplinarity and faculty placement differs across various university rank levels using quantile regression (see **Supplementary section 5** and **Fig. S7**). In all fields, higher Ph.D. interdisciplinarity was consistently associated with lower faculty placement prospects at the top-ranking universities. These results reinforce our earlier conclusion that Ph.D. interdisciplinarity is negatively associated with one's placement opportunities at top universities. Regarding heterogeneity across Ph.D. university ranks, across fields (including Math & Computing), the negative association between Ph.D. interdisciplinarity and faculty placement intensifies as the Ph.D. university rank increases and is most significant for top Ph.D. universities (see **Supplementary Fig. S8**). These results suggest that the disadvantages associated with high Ph.D. interdisciplinarity in faculty hiring are tied to the top tiers of Ph.D. universities.

**Interdisciplinarity matters for stayers in original fields**

Disciplinary transitions also occur during Ph.D. to faculty placement and are strongly associated with placement prospects[33]. We examined how the association between interdisciplinarity and faculty placement varies across three distinct placement types in disciplinary transition (see **Supplementary section 4**): (1) faculty who remain within their original Ph.D. field (same-field stayers), (2) those who transition to closely related fields (adjacent-field movers), and (3) those who make a significant shift to distant fields (distant-field movers). Our results indicate that across fields, the distant-field movers exhibit the highest levels of Ph.D. interdisciplinarity compared to other groups (see **Supplementary Fig. S9a**). Higher Ph.D. interdisciplinarity is positively related to a greater proportion of cross-disciplinary hires across all universities (see **Supplementary Fig. S9b**). However, the portion of movers (including both adjacent-field and distant-field movers) among faculty is higher in top universities (29.6% for top 5% and 27.4% for top 10% universities) compared to the overall sample (22.0%). The trend is consistent across fields, suggesting that top universities are more likely to hire faculty whose Ph.D. training occurred outside of their placement field (see **Fig. 2a**). With other variables controlled, distant-field movers are more likely to be hired by top universities compared to same-field stayers, which holds across fields and various thresholds for defining top universities (see **Supplementary Table S2**). This suggested the need for further investigation into the nuanced role of interdisciplinarity in the faculty hiring process.

**Fig. 2. Same-field stayers drive the negative association between Ph.D. interdisciplinarity and placement. a.** Share of faculty in the sample by field placement types in the top 10% and all universities. **b.** Marginal probability changes of placement types (non-top universities, same/adjacent/distant-field moves at top 10% universities) over a 1 standard deviation (SD) increase in Ph.D. interdisciplinarity, measured by Rao-Stirling index (*RS*). Marginal probabilities were estimated by the average marginal effects from multinomial logistic regressions with a full set of control variables. Error bars represent 95% confidence intervals. See **Fig. S5b-d** for results using other cutoffs (5%, 15%, 20%) of top university rank percentiles.

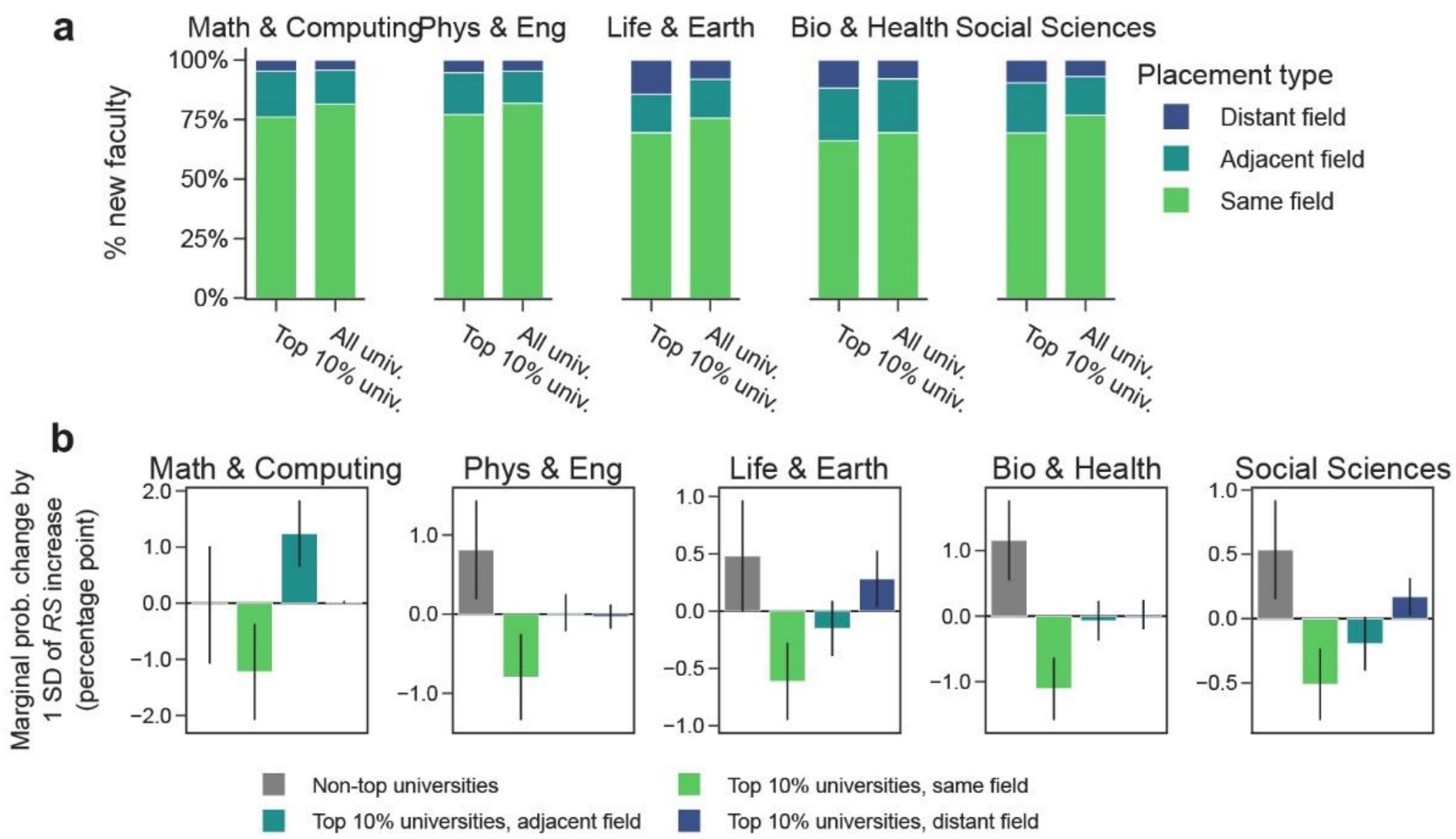

To clarify the role of Ph.D. interdisciplinarity in disciplinary transition and faculty placement, we conducted a multinomial logistic regression to analyze its relationship with university rank across different placement types. Defining the top 10% of universities as the "top", we observed that increases in Ph.D. interdisciplinarity are associated with higher probabilities of placement at non-top universities (i.e., those ranked below the top 10%) (see **Fig. 2b**). A 1 SD increase in Ph.D. interdisciplinarity is associated with a modest but consistent decrease in the probability of being hired at a top 10% university as a same-field stayer between 0.51 (Social Sciences, 95% CI [-0.79, -0.23], $p$<0.001) and 1.23 (Math & Computing, 95% CI [-2.08, -0.37], $p$=0.005) percentage points. These marginal effects can be meaningful in context: In Biology & Health, the share of our sample to be placed at the top 10% universities is 13%. Assuming 13% as a baseline probability, a 1 SD increase in interdisciplinarity can reduce a same-field candidate's chance by 8.5% relatively (i.e., shifting from 13% to 11.89%), and when aggregated over large applicant pools, such systematic shifts alter the composition of hires.

In contrast, among faculty hired into different fields (movers, including both adjacent-field and distant-field movers) at top universities, Ph.D. interdisciplinarity has smaller or even positive effects on placement probability. These trends remain robust under alternative rank thresholds for top universities (see **Supplementary Fig. S5b-d**) and raw Ph.D. interdisciplinarity (see **Supplementary Fig. S6b**). Additional logistic regression analysis disaggregated by placement type (see **Supplementary Table S3**) confirms that the overall negative association between Ph.D. interdisciplinarity and placement at top universities is primarily driven by the same-field stayers.

Together, these patterns reflect a two-sided selection. On the supply side, individuals who cross fields ("movers") tend to exhibit broader, more interdisciplinary doctoral training, enabling cross-field placement. On the demand side, top universities may hire a larger share of movers but favor boundary spanners whose breadth is instrumentally valuable, while also appointing many highly specialized stayers. This mix yields a higher fraction of movers yet a lower average level of measured interdisciplinarity among top hires, even as distant-field movers remain more likely to be placed at top institutions.

**Top universities prioritize research alignment in faculty recruitment**

Having established the role of Ph.D. interdisciplinarity in faculty placement, we now examine alignment with the research profile of the existing faculty. We aim to determine whether universities are more likely to recruit candidates whose research aligns with their strengths, reinforcing prevailing disciplinary or interdisciplinary trajectories. We hypothesize that a department's established interdisciplinarity level of its current faculty may shape hiring preferences regarding the interdisciplinarity of new hires. To test the hypothesis, we analyzed the average research interdisciplinarity of existing faculty (excluding new hires in our sample) across universities of varying ranks on a year-by-year basis. All these faculty included in this analysis were affiliated with the same departments and subfields as the new hires in our dataset, yielding a total sample of 193,979 faculty members. We find that faculty at higher-ranked universities (above the 90th percentile) generally exhibit lower research interdisciplinarity. This trend is particularly pronounced in Biology & Health and Social Sciences, where the average interdisciplinarity decreases by 0.06 (0.42 to 0.36) and 0.05 (0.40 to 0.35) from the $90^{th}$ to the $100^{th}$ percentile, respectively (see **Fig. 3a**). Furthermore, we observe a statistically significant positive correlation between the average interdisciplinarity of existing faculty and that of the new hires (see **Fig. 3b**). Notably, the correlation is relatively strong in Biology & Health (Pearson $r$=0.32, $p$<0.001) and Social Sciences ($r$=0.37, $p$<0.001). This suggests that the research environment of a department, particularly its orientation toward disciplinary versus interdisciplinary work, may be associated with hiring preferences.

**Fig. 3. Interdisciplinarity and research alignment with placement universities' faculty. a.** Existing faculty's interdisciplinarity across their university rank percentiles. Existing faculty were extracted year by year (2011-2020) and have the same university-subfield pairs as our sample. Their interdisciplinarity is calculated based on their past five years' publications. A gray line represents one of the extraction years. The red line shows the average of all gray lines. **b.** Positive correlations between a new faculty member's Ph.D. interdisciplinarity and their placement university's existing faculty's research interdisciplinarity**.** Existing faculty in the same department of a university as newly hired faculty in the same year were included in this analysis. Each dot represents a newly hired faculty member, colored by the rank percentile of the placement university. The slope ($\beta$), Pearson's correlation coefficient ($r$), and p values ($p$) are noted. **c.** The deviation of newly hired faculty's Ph.D. research from existing faculty's research in the same subfield, by placement university rank percentile and Ph.D. interdisciplinarity. The Ph.D. interdisciplinarity level is binned by its standard deviation from the field's mean (see **Supplementary Fig. S10** for other binning methods). The placement university rank percentile is binned into ten equal intervals. The color of each cell indicates the average deviation (see **Data and Methods**) for the corresponding faculty and university groups.

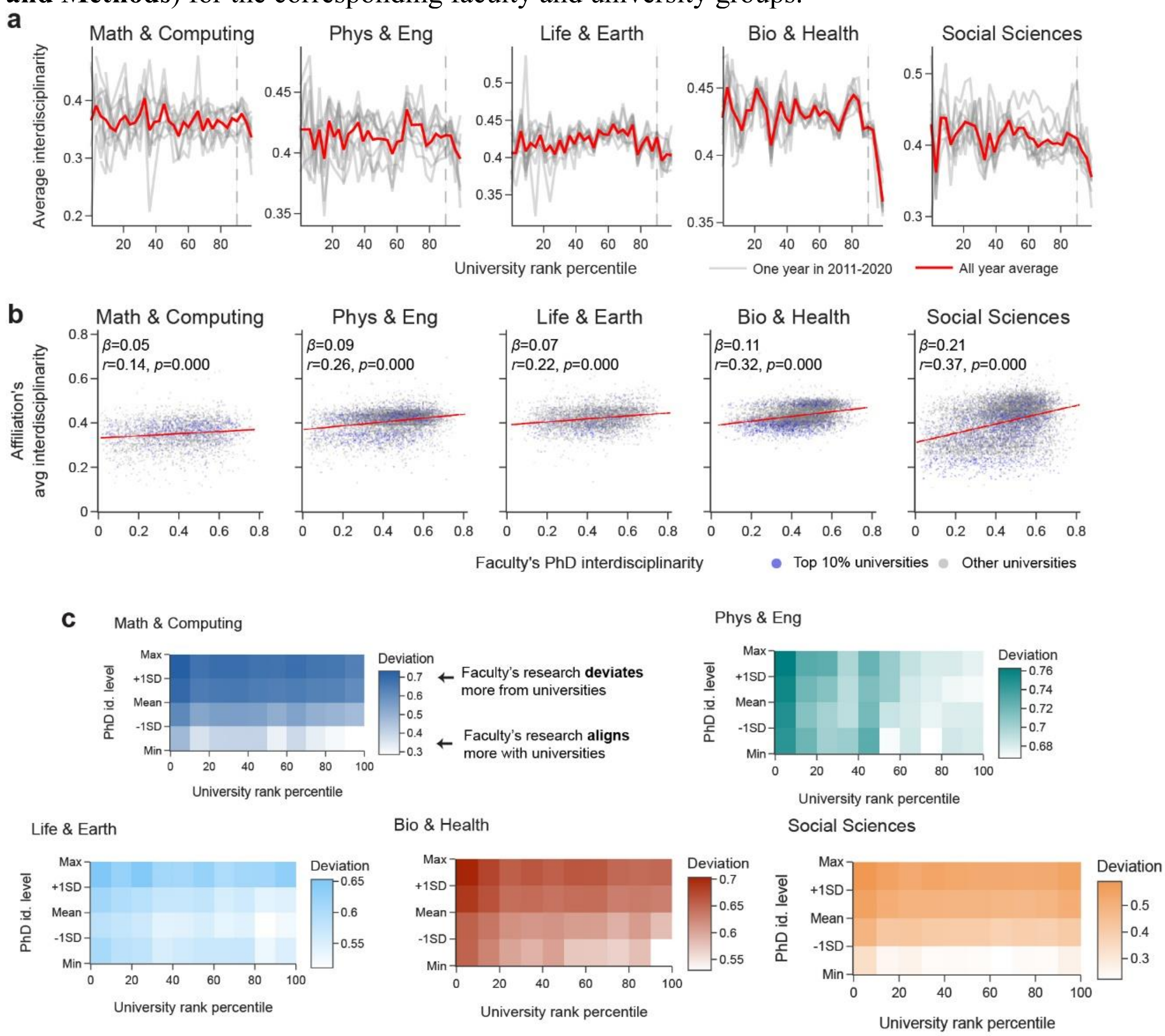

Beyond differences in interdisciplinarity, research alignment, or the degree to which a new hire's work aligns with the academic focus of existing faculty, may shape hiring decisions[35]. We measured the deviation between a new hire's Ph.D. publications and the research output of existing faculty at their placement universities (see **Data and Methods**). As shown in **Fig. 3c**, the rank of one's placement university is associated with research alignment: newly hired faculty with lower Ph.D. interdisciplinarity exhibit smaller deviations from the research profiles of existing faculty, a pattern prominent in higher-ranked universities. In contrast, newly hired faculty with higher Ph.D. interdisciplinarity consistently show larger deviations from the research profiles of faculty at placement universities, irrespective of university rank. This finding suggests that faculty with lower Ph.D. interdisciplinarity are more likely to align with the academic priorities of top universities, which may help improve their placement prospects.

**High interdisciplinarity is associated with gender imbalance in faculty placement**

The preference of top universities for new hires with lower interdisciplinarity may (un)intentionally create disparities in hiring opportunities for candidates whose academic profiles deviate from these expectations. Among such candidates, women may be disproportionately affected due to their tendency, on average, to pursue more interdisciplinary research compared to men[36–38]. To better understand the potential implications of these hiring preferences, we examine how gender and interdisciplinarity intersect in shaping faculty placement at top universities. Supporting previous findings[36–38], our findings indicate that among all the newly hired faculty in our analytical sample, the Ph.D. interdisciplinarity of women faculty is significantly higher than that of men (women 0.449, 95% CI [0.446, 0.451]; men 0.413, 95% CI [0.411, 0.415], Welch $t$-test $p<0.001$). This gender gap in interdisciplinarity holds in four out of the five fields examined (see **Fig. 4a**) and remains consistent across different faculty placement university ranks (see **Fig. 4b**), except Physics & Engineering ($p$=0.090). These results raise concerns that the hiring preferences of top universities favoring lower interdisciplinarity may unintentionally contribute to gender disparities in faculty placement by disadvantaging candidates whose research spans broader disciplinary boundaries.

**Fig. 4. Gendered Ph.D. interdisciplinarity in faculty placement. a.** Average interdisciplinarity by field and gender. Asterisks indicate the significance levels of gender differences based on Welch *t*-tests. **b.** Ph.D. interdisciplinarity versus faculty university rank percentile by fields and faculty gender. Trend lines were obtained using LOWESS smoothing. Shaded areas show 95% bootstrap confidence intervals. **c.** Logistic regression odds ratios comparing women to men in placement at top 10% universities, before and after propensity score matching (PSM) on faculty covariates. An odds ratio above 1 indicates that women were more likely than men to be placed in the top 10% of universities. "w/ PhD id." refers to matching on all covariates, including Ph.D. interdisciplinarity, while "w/o PhD id." excludes Ph.D. interdisciplinarity in the matching covariates. *** $p < 0.001$, ** $p < 0.01$, * $p < 0.05$.

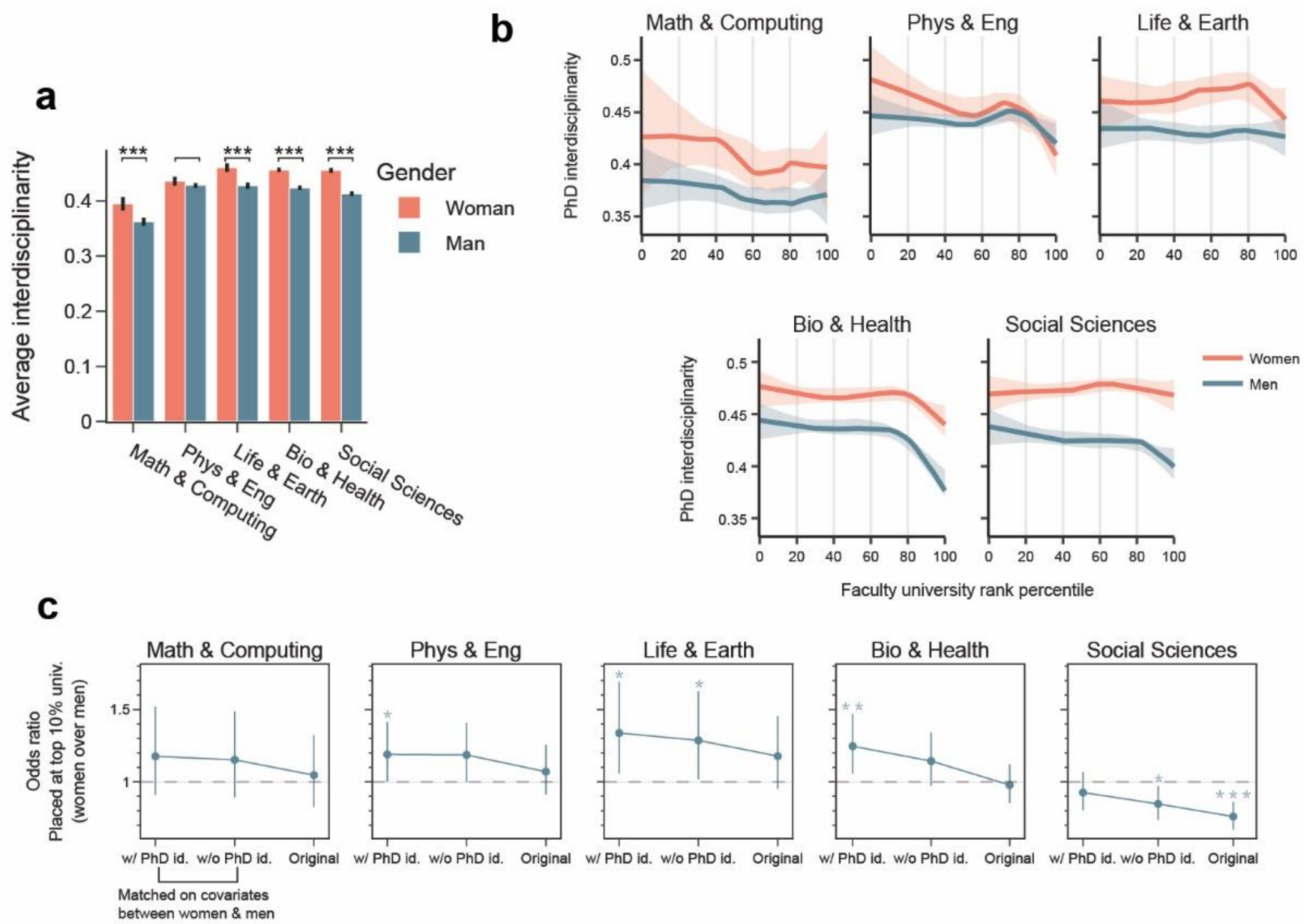

To isolate the effect of Ph.D. interdisciplinarity on faculty placement, we followed previous practices[39,40] and applied propensity score matching (PSM) within each field to construct two benchmark matched samples. In one specification, women and men are matched on the full set of control variables (excluding gender); in the other, Ph.D. interdisciplinarity is additionally included in the matching. This approach ensures that each matched sample balances observed covariates between women and men new hires, with the second sample also balancing their Ph.D. interdisciplinarity. Using the top 10% of universities as the definition of "top," our analysis reveals that in the PSM specification with Ph.D. interdisciplinarity, women were significantly more likely than men to be placed at top universities in Physics & Engineering (OR=1.190, 95% CI [1.002, 1.412], $p$=0.047), Life & Earth (OR=1.338, 95% CI [1.057, 1.691], $p$=0.015), and Biology & Health (OR=1.246, 95% CI [1.059, 1.467], $p$=0.008); in Social Sciences the difference is not statistically significant (OR=0.926, 95% CI [0.804, 1.066], $p$=0.287) (see **Fig. 4c**). In contrast, in the PSM specification without Ph.D. interdisciplinarity and the unmatched sample, the estimated likelihood of women being placed at top universities is lower, with women in Social Sciences at a disadvantage (OR=0.760, 95% CI [0.671, 0.860], $p$<0.001). In that field's unmatched sample, women comprised 48.3% of all new hires but only 37.1% of hires at top 5% universities (see **Supplementary Fig. S11**). In summary, the disadvantage associated with higher Ph.D. interdisciplinarity in faculty hiring at top universities disproportionately affects women, who exhibit higher levels of Ph.D. interdisciplinarity on average.

**Lower interdisciplinarity at top universities may hinder long-term knowledge production**

Beyond the implications for gender representation in academia, the preference for lower Ph.D. interdisciplinarity of new faculty hires at top universities raises a critical question: What are the potential long-term consequences for knowledge production? While disciplinary depth is valuable, an overemphasis on research alignment and a potential aversion to interdisciplinary approaches may hinder the creativity and intellectual cross-pollination that drives scientific breakthroughs[26].

We examined the potential long-term implications of top universities' hiring preference for candidates with lower Ph.D. interdisciplinarity, especially examining whether this preference might hinder knowledge advancement by limiting the inclusion of interdisciplinarity scholars. To assess this, we analyzed the publication records of faculty over a 10-year period following the completion of their Ph.D. degree, The results show that faculty with a higher level of Ph.D. interdisciplinarity outperform their less interdisciplinary peers in research productivity over the long term, especially at top universities (see **Fig. 5a** and **Supplementary Fig. S12**). Cumulatively, in Biology & Health at the top 10% of universities, for example, the new faculty members with above-mean Ph.D. interdisciplinarity on average published 51.3% more than those with below-mean Ph.D. interdisciplinarity within five years after Ph.D. graduation (mean 11.85±9.77 vs 7.83±6.34, median 10 vs 6, Mann-Whiteny *U* test $p$<0.001), and 40.0% more within 10 years (mean 28.04±26.4 vs 20.02±18.48, median 22 vs 15, $p$<0.001). When placed at non-top-10% universities, the corresponding percentages are 26.0% (mean 9.35±8.22 vs 7.42±6.79, median 7 vs 6, $p$<0.001) and 15.4% (mean 19.79±18.62 vs 17.15±16.77, median 15 vs 13, $p$<0.001), respectively. Prioritizing short-term fit over interdisciplinary potential could, therefore, represent a missed opportunity for fostering more impactful and wide-ranging scientific contributions.

**Fig. 5. Long-term benefits of Ph.D. interdisciplinarity. a.** Annual productivity during Ph.D. training and 10 years after graduation at top 10% universities, categorized by field and Ph.D. interdisciplinarity level. The publication records (up to 2020) of faculty members in our sample were pooled together based on their Ph.D. graduation year (denoted as year 0 on the x-axis). The mean values of Ph.D. interdisciplinarity level were computed within fields. Trend lines were obtained using LOWESS smoothing. Shaded areas show 95% bootstrap confidence intervals. **b.** Average productivity changes per 1 standard deviation (SD) increase in Ph.D. interdisciplinarity by field and faculty university rank. Productivity changes were estimated by the average marginal effects from Poisson regression using the full set of control variables.

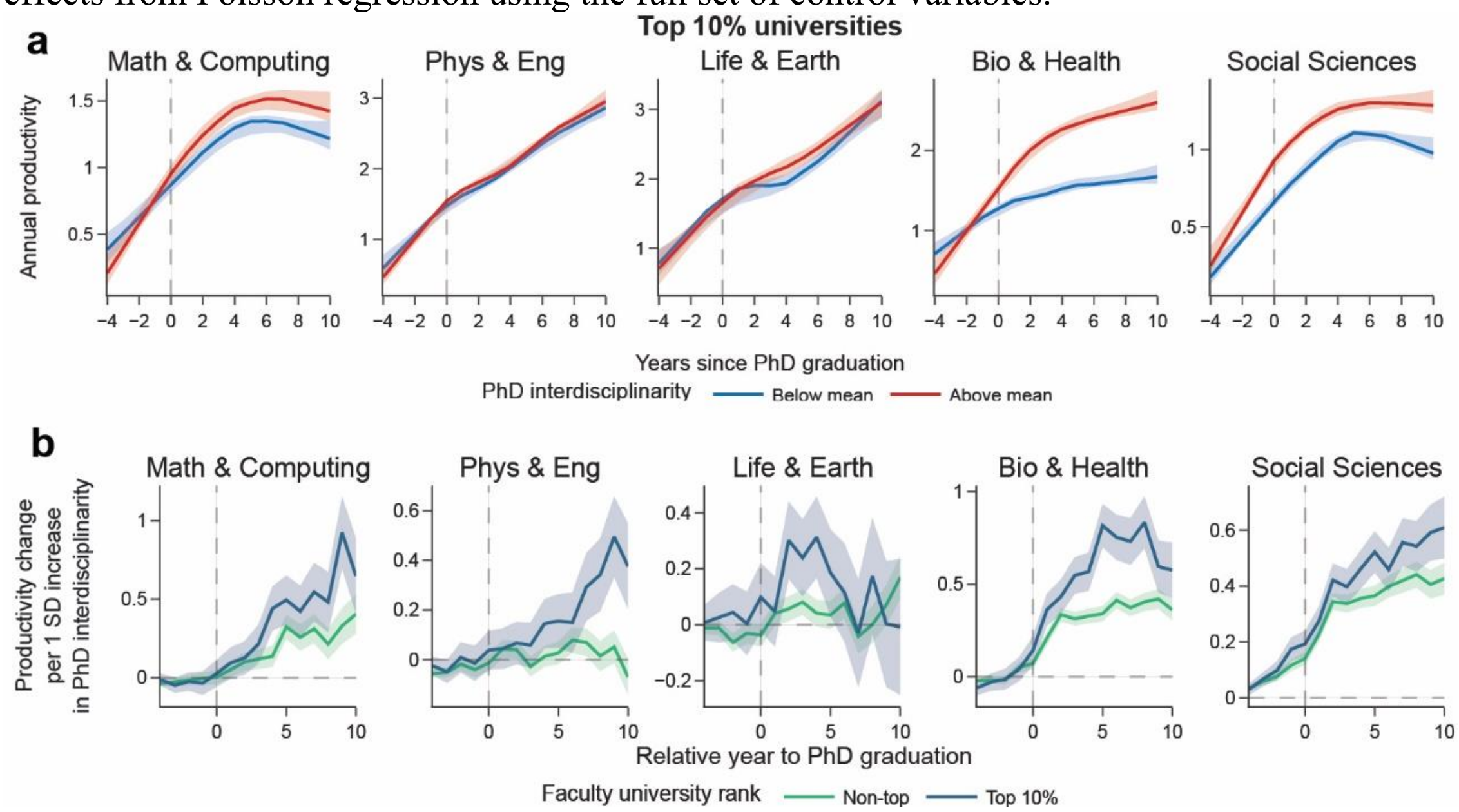

To assess whether this productivity advantage is associated with higher interdisciplinary varies across universities, we fitted Poisson regression models to estimate the average marginal effects of Ph.D. interdisciplinarity on future annual productivity (see **Data and Methods**). Among new hires at top universities, a 1 SD increase in Ph.D. interdisciplinarity is generally associated with greater long-term research productivity after graduation (see **Fig. 5b** and **Supplementary Fig. S13**). For instance, in Biology & Health at top 10% universities, a 1 SD increase in Ph.D. interdisciplinarity corresponds to an average of publishing 0.82 papers more in the fifth years after Ph.D. graduation (95% CI [0.70, 0.93], $p$<0.001). In Social Sciences, the 1 SD increase is associated with publishing 0.52 more papers in the same year (95% CI [0.45, 0.60], $p$<0.001). In contrast, at non-top universities, the corresponding increases are smaller: 0.34 papers in Biology & Health (95% CI [0.30, 0.38], $p$<0.001) and 0.36 papers in Social Sciences (95% CI [0.33, 0.40], $p$<0.001). Disciplinary patterns also vary. In Life & Earth sciences, productivity gains from Ph.D. interdisciplinarity peak in the fourth year after graduation and decline thereafter, while in other fields, the benefits tend to peak later. Faculty across different interdisciplinarity levels exhibit similar normalized citation impact across university ranks (see **Supplementary Fig. S13**). Nevertheless, by favoring lower-interdisciplinary hires, the top universities may be missing opportunities to increase their research output and contribute to knowledge.

## Discussion

This study reveals the paradoxical role of interdisciplinarity in Ph.D. training in faculty hiring at top universities. While interdisciplinarity is often celebrated for fostering innovation[1–3], scientists who leave the safe haven of their home discipline to explore the territory between fields are often punished rather than rewarded by the academic system[27]. In addition to previous findings that interdisciplinary Ph.D. graduates may be disadvantaged in retaining positions in academia[8,41], our findings suggest that faculty with higher Ph.D. interdisciplinarity are less likely to be placed at top universities across fields, except for Math & Computing (see **Fig. 1**). Despite the institutional rhetoric advocating interdisciplinary and the widespread promotion and adoption of interdisciplinarity as an institutional goal or strategy among leading research universities, the hiring pattern at top universities suggests persistent preferences for candidates with more traditional, discipline-aligned research trajectories.

The hiring disadvantage is particularly pronounced for faculty who obtained Ph.D. from top universities and "stayed" in their Ph.D. subfield when transitioning into faculty roles for all fields, including Math & Computing. In contrast, interdisciplinary faculty who move into faculty positions at a department with a different disciplinary focus experience fewer such penalties (see **Fig. 2**). This suggests that top universities may consider disciplinary coherence when recruiting faculty within the same field. This aligns with our analysis of the existing faculty's interdisciplinarity, showing those at top universities have lower research interdisciplinarity (see **Fig. 3a**). Previous research indicates that the disciplinary structure of US universities is a factor impeding faculty from expanding ongoing interdisciplinary research activities[42]. While top universities may benefit from disciplinary structures that foster cohesive scientific communities and allocate resources effectively to sustain high-quality, specialized research, it may also limit institutions' ability to integrate and sustain interdisciplinary perspectives[11,43]. Although universities have taken steps to promote interdisciplinary collaboration, such as implementing interdisciplinary cluster hiring, the collaboration remains largely confined within disciplinary boundaries[44].

A plausible explanation for this trend is that the new hires with lower Ph.D. interdisciplinarity are more likely to align with the established research priorities of top universities. Our analysis of new faculty's Ph.D. research in comparison to the research profiles of existing faculty reveals that candidates with lower Ph.D. interdisciplinarity are more likely to produce work that more closely matches the academic focus of top universities (see **Fig. 3c**). Interdisciplinary research has long been regarded as more difficult to evaluate[9,14,26,32]. This misalignment may bring cognitive challenges for existing faculty to appraise interdisciplinary work and/or accept such work as part of their domain[9,24]. Furthermore, interdisciplinary scholars may face greater challenges in establishing themselves within professional communities and a systemic bias favoring established disciplines in funding decisions[27]. Interdisciplinary faculty members are perceived to be punished during the tenure process for venturing beyond their home disciplines[42], and researchers may become more disciplinary over time in response to the pressure[8].

Our findings also highlight the role of Ph.D. interdisciplinarity in gender disparities in faculty hiring at top universities. Across fields and university ranks, women consistently exhibit higher levels of Ph.D. interdisciplinarity compared to their male counterparts across fields and university ranks (see **Fig. 4a, b**), which aligns with previous studies[36–38]. Given top universities' preference for lower Ph.D. interdisciplinarity, this hiring bias may disproportionately disadvantage women, particularly in fields such as Social Sciences, where gender imbalances in hiring are pronounced at top universities (see **Supplementary Fig. S11**). However, after balancing interdisciplinarity using propensity score matching, women exhibited a higher likelihood of securing positions at top universities (see **Fig. 4c**). This suggests that Ph.D. interdisciplinarity serves as a key factor influencing gender disparities in hiring and that accounting for it can reduce or even reverse observed gender gaps. Addressing institutional biases against interdisciplinarity in hiring practices could therefore be an essential strategy for fostering gender equality in the academic workforce.

Beyond its implications for faculty diversity, the preference for lower Ph.D. interdisciplinarity among top universities has long-term consequences for research productivity. Our analysis reveals that faculty with higher levels of Ph.D. interdisciplinarity tend to outperform their less interdisciplinary peers in publication output over the first decade after Ph.D. graduation, with disciplinary variations (see **Fig. 5a**). This finding extends previous research linking interdisciplinarity to increased productivity[9], highlighting that early Ph.D.-period interdisciplinarity can contribute to long-term benefits. Interestingly, this productivity advantage is even more pronounced at top universities, where faculty with higher Ph.D. interdisciplinarity experience the largest marginal increases in publication output (see **Fig. 5b**). However, despite the productivity advantage, top universities' reluctance to hire interdisciplinary Ph.D. graduates may lead to missed opportunities for enhancing scientific output. These findings call for a reevaluation of hiring practices that systematically disadvantage interdisciplinary scholars, as such biases may contribute to the high dropout rate of interdisciplinary scholars from academia and ultimately undermine the research ecosystem[8].

In sum, our findings reveal that top universities' hiring preferences for lower Ph.D. interdisciplinarity not only disadvantage interdisciplinary scholars, particularly women, but also pose broader challenges to scientific productivity. Addressing these barriers through more inclusive hiring practices could help universities better align with their stated commitments to interdisciplinarity, while also fostering a more diverse, equitable, and dynamic research environment.

Several limitations must be acknowledged. Our analysis focuses exclusively on Ph.D. graduates who secured faculty positions at US institutions, omitting a substantial proportion of Ph.D. graduates across the interdisciplinarity spectrum who did not get faculty positions. Therefore, our findings may not apply to individuals who choose different career paths, e.g., industry and non-academic roles, and who left the US. Our reliance on bibliographic and faculty rosters data may not fully capture informal factors of Ph.D. training and faculty hiring, such as mentorship quality, professional networks, campus and departmental culture, and decision-making process. Our measurement of interdisciplinarity essentially reflects the diversity of the knowledge base upon which one's research is built, which may not reflect the complete spectrum of interdisciplinary engagement. Longitudinal analyses or survey-based approaches can better track career outcomes over time and explore how interdisciplinary scholars navigate mid-career transitions. Additionally, examining interventions that support interdisciplinary scholars, such as interdisciplinary clusters, funding programs, or dual appointments, could provide actionable strategies for universities seeking to foster innovation and equity in academic hiring.

## Data and Methods

**Data.** Our analytical sample was sourced from the U.S. higher education university faculty rosters dataset by the Academic Analytics Research Center (AARC). This dataset, as used in previous research[45], includes tenure-track faculty employment records from the U.S. Ph.D.-granting universities from 2011 to 2020, tracking the career trajectories of 314,141 tenure-track faculty members from 393 universities. For each faculty member, the dataset provides their basic information such as name, Ph.D. universities, and Ph.D.-degree-awarded year. For each year in the time window, the dataset includes each faculty member's affiliation, department, and job title. Each professor has been linked to their publication history from 2006 to 2020, with digital object identifiers (DOIs) attached to papers.

We supplemented this dataset by linking it to the ProQuest Dissertation & Thesis (ProQuest) database, updated to 2020. We matched individual faculty members in AARC to ProQuest using full name, Ph.D. universities, and Ph.D. completion year (corresponding to the dissertation year in ProQuest). Faculty members with multiple matches were excluded. We limited the sample to individuals whose Ph.D. degrees were awarded between 2005 and 2018 to ensure a sufficient sample size in each year and the capability to capture initial faculty placement. We used each individual's earliest record of faculty affiliation and department name in the AARC dataset to represent their initial faculty placement. We assigned faculty to one of 24 subfields and six broad fields based on the AARC's classification of faculty departments. We also identified their Ph.D. subfields and fields based on ProQuest departments and imputed the missing values from dissertation titles using machine learning (see **Supplementary section 1**). We removed faculty whose Ph.D. was in the Arts and Humanities field due to the low coverage of Arts and Humanities publications in bibliographic databases.

Given that many Ph.D. graduates pursue one or multiple postdoctoral positions before securing a faculty job, and it may take several years of applying before securing a faculty position, we included individuals in our sample whose earliest record is within six years of their Ph.D. graduation year, keeping 49,508 faculty members at this stage. We estimated that among faculty who had held their Ph.D. degree for six years, approximately 90.9% remained at the same institution as their initial faculty placement (see **Supplementary Fig. S14**). Therefore, our method should most accurately reflect individuals' initial faculty placement. Considering the potential impact of transitional non-faculty positions, such as postdocs, between Ph.D. graduation and initial faculty placement, we selected a subsample of 111,71 faculty members who graduated after 2009 and had the earliest faculty records no more than two years after their Ph.D. graduation. The test results are qualitatively consistent with the previous results in **Fig. 1** and **Fig. 2** (see **Supplementary Fig. S15**).

We enriched the publication information of faculty recorded by AARC with Clarivate's Web of Science (WoS) for a comprehensive publication history for the studied faculty sample. Specifically, we used the DOIs of faculty journal and conference proceeding publications recorded by AARC to find additional bibliographic information, including the references, citations, and collaborator information of each DOI. To remove papers by oversized teams where individual contributions might be exceedingly diluted, we removed 1.9% of papers with more than 10 authors. We also removed 4.9% of papers with fewer than five references in WoS in order to calculate interdisciplinarity (see below). Considering the delay between manuscript submission and final publishing, we collected an individual's Ph.D.-stage papers within a six-year window, from four years before graduation (year -4) to one year after (year 1). We kept those with at least one paper during this period to calculate their Ph.D. interdisciplinarity, which accounts for 69.8% of all individuals. Among the rest, we removed 3% of outlier individuals whose productivity during Ph.D. training is higher than 20 papers and 3.6% who could not be assigned a binary gender (see **Supplementary section 3**) and/or were awarded with non-US Ph.D. degrees. Our final analytical sample includes 31,775 faculty members, along with their 852,031 lifetime papers, including 166,251 papers completed during Ph.D. training.

**Paper and Ph.D. interdisciplinarity.** We measured a paper's interdisciplinarity by the disciplinary diversity of its references, also known as knowledge-base interdisciplinarity, which represents the interdisciplinarity of the knowledge base upon which the paper is built to support its claims[7,46]. We used the Rao-Stirling index (*RS*) to calculate the interdisciplinarity of each paper[2]. Specifically, we extracted a year $t$-published paper $s$'s reference list and obtained each reference's discipline according to our journal discipline classification system, which categorizes WoS journals into 144 low-level disciplines[47]. If the paper has at least five references with discipline information, we continue to calculate its interdisciplinarity as

$$IDR_s = 1 - \sum_{ij} S_{ijt} P_{is} P_{js}$$

where $P_{is}$ is the proportion of references categorized in discipline $i$ for paper $s$, and $S_{ijt}$ is the similarity score between discipline $i$ and discipline $j$ for year $t$. To calculate this similarity score, we estimated the overall knowledge base structure within disciplines across all WoS papers and years. For discipline $i$ and year $t$, we built a 144-dimensional cumulative reference vector $V_{it}$ over all WoS discipline-$i$ papers published in year $t$, where each vector element is the share of references in one of the 144 disciplines out of the total reference count. After building vectors $V_{it}$ and $V_{jt}$ for discipline $i$ and $j$, their similarity score $S_{ijt}$ is defined as the cosine similarity between $V_{it}$ and $V_{jt}$. After computing all papers' interdisciplinarity in our data, for each faculty member, we calculate their Ph.D. interdisciplinarity as the median interdisciplinarity score of all the papers they published during their Ph.D. stage, as defined above[8]. We also tested the results using the Rao–Stirling index calculated from the pooled aggregate references of those papers and found similar results (see **Supplementary section 5** and **Fig. S16**).

**University rank by subfield.** We applied the SpringRank algorithm[48], a physically inspired model to infer hierarchical rankings of nodes in directed networks, over the AARC data within each subfield. Rankings based on the SpringRank algorithm follow the principle that interactions are more likely to occur between individuals with similar ranks. Based on the Ph.D. and faculty affiliations and subfields for all faculty in AARC data, we applied the SpringRank algorithm and generated a real-valued rank for each university-subfield pair. We converted the real value ranks into rank percentiles to facilitate the recognition of top universities. Therefore, in our analysis, "top 10% universities" refer to universities ranked in the 90th percentile or higher in the corresponding Ph.D. subfield. Note that because AARC datasets only contain US universities, non-US universities are not ranked in our analysis and thus excluded. See **Supplementary Table S4** for the number of universities and top universities in the rankings by subfield. We also tested the results, setting the top universities by four popular world university rankings (see **Supplementary section 5** and **Fig. S17**).

**Regression analysis for faculty placement.** To analyze the relationship between Ph.D. interdisciplinarity and the likelihood of top-rank faculty placement, we estimated binary logistic regression models:

$$log\left(\frac{Pr(y=1)}{Pr(y=0)}\right) = \beta_0 + \beta_1 IDR + \mathbf{X}_{\gamma} + \epsilon$$

where $y$ is a binary indicator for placement in the top X% of universities (X set at 10% in **Fig. 1e** and other thresholds in **Supplementary Fig. S5**), $IDR$ represents standardized Ph.D. interdisciplinarity, and $\mathbf{X}_{\gamma}$ denotes control variables. As we standardized $IDR$ within fields, its coefficient $\beta_1$ can be interpreted as the association of a one-standard-deviation increase in the outcome. The full set of covariates includes Ph.D. university rank within the respective subfield, Ph.D. graduation year, personal features (gender, Ph.D. publication count, average normalized citations, unique collaborator count during Ph.D.), and advisor features (Ph.D. advisor's gender, advisor's 5-year publication count since five years before Ph.D. entry year, and advisor's seniority up to the Ph.D. entry year; see **Supplementary section 2**). Citations are normalized by the average citation counts within each discipline and publication year in WoS. The advisor's publication count reflects output in the five years prior to the Ph.D. entry year (i.e., years -9 to -5 relative to the Ph.D. graduation year). An advisor's seniority is defined as the interval between the advisor's first publication and their student's Ph.D. entry year, and then binned into five quintiles. The cutoff point at Ph.D. entry year is intended to exclude advisor-student collaborative papers during Ph.D. years.

**Research deviation between newly hired faculty and existing faculty.** To assess how a faculty member's research aligns with one university's research within the same subfield, we compared the cumulative reference compositions of the new faculty member and the university's existing faculty in that subfield. We conducted this comparison for every possible pair of faculty members and universities in our sample. Specifically, for each faculty member in our sample, we constructed a 144-dimensional cumulative reference vector based on the faculty member's Ph.D.-stage publications, where each vector element represents the relative share of references categorized into one of 144 disciplines. Next, for each faculty member, we identified the earliest year the faculty member appeared in the AARC dataset and then extracted all registered existing faculty at each university within departments of the same subfield as the faculty member. For each of these existing faculty members, we gathered their publications from the preceding five years to build individual cumulative reference vectors. In total, these vectors cover 238,114 existing faculty and their 2,682,607 publications. To represent the university's overall research focus in one subfield, we aggregated these individual vectors by university and subfield by averaging across the existing faculty. The research deviation was then quantified as the cosine distance between each faculty member's cumulative reference vector and the university-subfield-level reference vector.

**Propensity score matching**. To evaluate the gender difference in opportunities for top-rank faculty placement after accounting for Ph.D. interdisciplinarity, we performed propensity score matching (PSM) to match women and men faculty members by a set of confounding covariates. Following previous research using PSM to examine gender differences[39,40], we first estimated each faculty member's probability (propensity score) of being a woman given observed characteristics, i.e., the full set of control variables excluding gender. Then, we applied radius matching with a caliper of 0.1 to match the propensity scores and enhance the covariate similarity regarding the control variables between women and men faculty. This matching process was conducted twice: For the first time, using a full set of control variables excluding gender and Ph.D. interdisciplinarity, and for the second time, adding Ph.D. interdisciplinarity as an additional covariate. We estimated the post-matching gender effect on the faculty placement at the top 10% of universities using weighted logistic regression, which uses the faculty falling within the common support region and assigned with matching weights generated from PSM. **Supplementary Table S5** presents the results of covariate balance checks of all PSM specifications across fields. After estimation, we tested whether the log-odds differences in being in the top 10% of universities for women compared to men were significant. This analysis was supported by *psmatch2* package in Stata[49].

**Long-term research performance.** To track the long-term research performance of faculty members post-Ph.D., we analyzed publication records of our sample by aligning these records by years relative to Ph.D. graduation. We capped the analysis at ten years post-graduation to retain sufficient sample sizes for each year. We applied the following Poisson regression model:

$$\log E(n_{it}) = \beta_0 + \sum_{\tau=-4}^{10} \delta_\tau I(\tau = t) + \beta_1 IDR_i + \beta_2 Top_i + \sum_{\tau=-4}^{10} \gamma_\tau I(\tau = t) \cdot IDR_i + \sum_{\tau=-4}^{10} \gamma_\tau I(\tau = t) \cdot Top_i + \beta_3 IDR_i \cdot Top_i + \sum_{\tau=-4}^{10} \theta_\tau I(\tau = t) \cdot IDR_i \cdot Top_i + \mathbf{X}_{\boldsymbol{\gamma}}$$

where $n_{it}$ denotes the annual paper number for faculty $i$ in relative year $t$, and $Top_i$ is a binary variable that equals 1 if the faculty university is ranked above the top university threshold. Standard errors are clustered at the faculty member level. $IDR$ is similarly standardized within fields. From the Poisson regression model, we estimated the average marginal effects of Ph.D. interdisciplinarity on the annual number of publications for every career year, with $Top_i$ being fixed at 0 (non-top) and 1 (top) separately.

## Acknowledgments

We thank the Academic Analytics Research Center, ProQuest, and Observatoire des sciences et des technologies at the University of Quebec in Montreal for sharing data. We thank the Center for High Throughput Computing for computing resources[50]. This work is supported by the Institute for Humane Studies under grant no. IHS018809, the Wisconsin Alumni Research Foundation, and the Vilas Life Cycle Professorships.

## Author Contributions

XZ and CN designed research; XZ and CN performed research; XZ, XH, and CN analyzed data; and XZ, AP, XH, CS, and CN wrote the paper.

## Competing Interest Statement

The authors declare no competing interests.

## Supplementary Information

### 1. Field assignment

We assigned faculty to academic fields based on the AARC's original classification of faculty departments, which includes 187 granular taxonomies. From these taxonomies, we categorized each faculty member into one of 24 subfields and six broad fields: Math & Computing, Physics & Engineering, Life & Earth, Biology & Health, Social Sciences, and Humanities. Humanities were removed from the main analysis due to the low coverage of Humanities publications in the Web of Science[51]. The same method was applied to determine the Ph.D. subfields for faculty members whose Ph.D. department information was available.

Among all faculty members that we matched ProQuest and AARC datasets who graduated after 2005, 78.4% of Ph.D. department data was missing in our snapshot of ProQuest. To address this limitation, we trained machine learning models on 272,956 ProQuest dissertations from 2005 to 2018 that were not included in our AARC-ProQuest faculty matching results and had available Ph.D. department names for the dissertation authors. Using their department names, we mapped these dissertations to 24 unique subfields based on AARC's taxonomy classification of the same department names. The mapped subfields were used as the labels of the dissertation training data for the models. For predictive features, we utilized the subject categories recorded by ProQuest. A dissertation author selects the field(s) of research associated with their work from a list of subject categories when submitting their thesis to ProQuest, and the classification system for subject categories has been relatively stable in recent decades[52]. In our training data, each dissertation has at least one subject category and may have multiple subject categories, which amount to a total of 444 distinct subject categories.

Using the extracted dissertation training data, we trained models including neural network, logistic regression, support vector machine, random forest, and XGBoost based on the predictive features of subject categories and outcome classification categories. Each model's hyperparameters were optimized using 5-fold cross-validation. Then, we utilized 9,964 dissertations by our matched AARC-ProQuest faculty with available Ph.D. department information. We randomly used 70% of these dissertations sampled by subfield to ensemble the predicted probabilities from each model by training a logistic regression model. To evaluate the performance, we tested it on the remaining 30% and achieved 86% accuracy and an F1 score of 0.86 (see **Table S6** for prediction performance). We then apply the trained models to predict missing Ph.D. subfields for the rest of our sample (see **Table S1** for the distribution of faculty members by faculty and Ph.D. field and subfield).

**2. Advisor identification and publications**

Whenever available, ProQuest includes the full name of each Ph.D. student's advisor, offering a possible clue for tracing the advisor's career and research trajectory. We identified the faculty members' advisors' publication records by two methods: First, we matched advisors' names within the faculty's publication collaborator list and then linked the collaborators to our disambiguated WoS author identifiers. Using the identifiers, we extracted their WoS-indexed publications. Second, we searched for these advisors' full names and affiliations (based on the faculty's Ph.D. universities) from AARC. Once matched, we were able to retrieve their publication DOIs recorded by AARC, which were then further matched with WoS to supplement the publication records. The two publication records were finally pooled and deduplicated. Advisor seniority was determined by calculating the time from their earliest publication to their student's Ph.D. start year (four years prior to the graduation year), and then dividing into five quintiles. For faculty who had multiple advisors listed in ProQuest, we kept their first-listed advisors. We removed faculty members whose gender could be assigned and had no publication records before the Ph.D. start year (i.e., seniority of 0 years). We retrieved Ph.D. advisors for 29,096 (91.6%) of faculty members in our sample.

**3. Gender assignment**

We used three approaches to infer and impute faculty members' binary gender. First, we leveraged an online teaching evaluation dataset from RateMyProfessors.com, consisting of 919,750 professors and instructors, to extract gender signals from name titles (e.g., Mr. and Ms.) and student comments (mentioning gender pronouns such as he, she, his, him, her, and hers)[53]. Only consistent signals were used at this stage, and ambiguous or conflicting cases were deferred to the next step. These gender signals used independently by professors and students' comments provide convergent evidence about an instructor's perceived binary gender. Second, for the remaining individuals, we incorporated precomputed gender labels from AARC, which were generated via NamSor's name-based gender inference algorithm[54]. Third, for any remaining faculty members, we applied an algorithm that aims to infer Web of Science author gender using U.S. census data and country-specific name lists[55,56]. We drew a random sample of 8,000 professors to compare the three gender-assignment methods; they concurred on a binary gender in 90.4% of instances with successful binary gender identification. We applied the same procedures to assign genders to their Ph.D. advisors. Finally, we identified 96.8% of faculty members' and 96.4% of Ph.D. advisors' binary genders in our sample. We admit the limitation that our gender assignment pipeline cannot distinguish non-binary gender and may have errors in the results regarding these minority groups.

**4. Disciplinary distance in Ph.D.-faculty transition**

We calculated the disciplinary distance across department subfields to determine the placement types ("same-field stayers," "adjacent-field movers," and "distant-field movers") in Ph.D.-faculty transition by constructing a directed network of Ph.D.-to-faculty subfield mobility. This subfield mobility network uses all 58,278 faculty members who graduated after 2005 and were identified with Ph.D. and faculty subfields (see **Fig. S18a**). Nodes in the network represent the subfields, and edges represent the disciplinary movement of faculty from one subfield to another when transitioning from Ph.D. to faculty. Edge weights were initially based on the number of faculty members who switched from a source subfield to a target subfield. However, these weights can be influenced by the relative sizes of the source and target fields. To address this effect, we performed 100 random shuffles of faculty subfields across the faculty included in the entire network, including faculty who remained in their original Ph.D. subfields. For each shuffle, we reconstructed the network following the same procedure. The edge weight from subfield A to subfield B in the original network was then normalized by dividing it by the average edge weight from subfield A to subfield B across the 100 shuffled networks. We defined the distance between two fields as the reciprocal of the normalized edge weight between them. To identify the distance between these subfields, we examined the cumulative distribution of faculty members who switched their subfields. We observed that for most subfields, the distribution trend lines exhibit an inflection point around the 80th percentile of all faculty (see **Fig. S18b**). Accordingly, we define subfields with distances below this point as adjacent fields, while those above are distant fields.

**5. Robustness check**

We conducted several robustness checks to ensure that our core findings about Ph.D. interdisciplinarity and faculty placement are not driven by particular measurement choices, model specifications, ranking definitions, omitted covariates, or sample selection.

**(1) Interdisciplinarity measurement**. In the main results, we standardized the measure of the Rao-Stirling index by subtracting the mean from each value and then dividing by the standard deviation to facilitate the interpretation of regression results. As a robustness check to confirm that standardization does not drive the results, we also replicated the regressions using untransformed Ph.D. interdisciplinarity within each regression sample and found consistent results (see **Fig. S6**).

**(2) Alternative construction of Ph.D. interdisciplinarity.** Instead of calculating each individual paper's interdisciplinarity and then taking the median of paper-level scores, we pooled all Ph.D. papers' distinct references (i.e., a set of references with no duplicates) and calculated an individual's Ph.D. interdisciplinarity based on the disciplinary distribution of the pooled references. The results align with our conclusion that faculty at top universities have lower Ph.D. interdisciplinarity (see **Fig. S16**).

**(3) University rank definition.** To ensure that the results are insensitive to the choice of ranking method, we substitute our data-driven university rank percentiles with external four major world university ranking systems: the Academic Ranking of World Universities (ARWU), QS World University Rankings (QS), Times Higher Education World University Rankings (Times), and U.S. News & World Report Best Global Universities Rankings (US News). For convenience, we defined the top U.S. universities as those ranked within the top 20 nationwide by a world university ranking in 2020, irrespective of field. As illustrated in **Fig. S17**, we applied the same regression analysis and observed consistent results in **Fig. 1** and **Fig. 2**.

**(4) Model specification across the rank distribution.** In addition to the binary logistic regression, we employed quantile regression to examine whether the association between interdisciplinarity and placement varies across the full spectrum of faculty university ranks. Quantile regression allows us to observe how interdisciplinarity might impact placements from lower to higher-ranking universities. The model specification is

$$Q_y(\tau) = \beta_{0,\tau} + \beta_{1,\tau} IDR + \mathbf{X}_{\boldsymbol{\gamma}}$$

where $Q_y(\tau)$ represents the conditional quantile of the faculty university rank at quantile $\tau$. $\mathbf{X}_{\boldsymbol{\gamma}}$ denotes a full set of covariates, including Ph.D. university rank within the subfield, Ph.D. graduation year, gender, Ph.D. publication count, average normalized citations, unique collaborator count during Ph.D., Ph.D. advisor's gender, advisor's 5-year publication count five years before Ph.D. start, and advisor's seniority quintile up to the Ph.D. start. The coefficients from the regression models are summarized in **Fig. S7**. We observed that in all fields, higher Ph.D. interdisciplinarity was consistently associated with lower faculty placement prospects at the top-ranked universities (red coefficient points and error bars), which is insignificant in the lower-ranked universities (gray coefficient points and error bars). This indicates that the negative association between Ph.D. interdisciplinarity and university ranking mainly exists among top faculty universities, which aligns with our conclusions.

**(5) Controlling the collaboration network structure**. We tested the full set of controls in the logistic regressions shown in **Fig. 1** with measures of each newly hired faculty member's Ph.D.-stage collaboration network structure. For each individual, we built an egocentric collaboration network using publications from four years before Ph.D. graduation through one year after; nodes comprise the focal faculty and their collaborators, and each individual pair's edge is weighted by the number of their co-authored papers in that time window. We then computed four centrality metrics for the focal faculty: degree, betweenness, closeness, and eigenvector centrality. To ensure these metrics were reliably estimated, we restricted the sample to faculty with at least three distinct collaborators, resulting in 18,431 faculty members. Including each centrality measure as additional control did not alter the substantive results reported in **Fig. 1** (see **Table S7**).

**(6) Sample restriction for transitional positions.** We also examined the results by limiting the sample to faculty whose initial appointments occurred within two years of Ph.D. graduation to mitigate confounding from postdoctoral or other interim positions. This is a subsample of 11,171 faculty members who graduated after 2009, as our faculty rosters data's earliest record is from 2011. The results from this restricted sample are qualitatively consistent with the primary findings in **Fig. 1** and **Fig. 2** (see **Supplementary Fig. S15**).

**Fig. S1. Faculty's Ph.D. interdisciplinarity and university rank. a.** The flow of faculty affiliations from their Ph.D. universities to faculty universities. Percentages represent the distribution of Ph.D. graduates across the rank percentiles of their Ph.D. granting institutions and the corresponding distribution of their faculty positions across university rank percentiles, illustrating the flow of faculty placement. **b.** Distribution of Ph.D. interdisciplinarity by fields. Shaded areas show the overall distribution across all fields. Means and standard deviations (SD) of Ph.D. interdisciplinarity for each field are noted. **c.** Increasing trends in the average Ph.D. interdisciplinarity by graduation year. Trend lines were generated using LOWESS smoothing, with shaded areas representing 95% bootstrap confidence intervals.

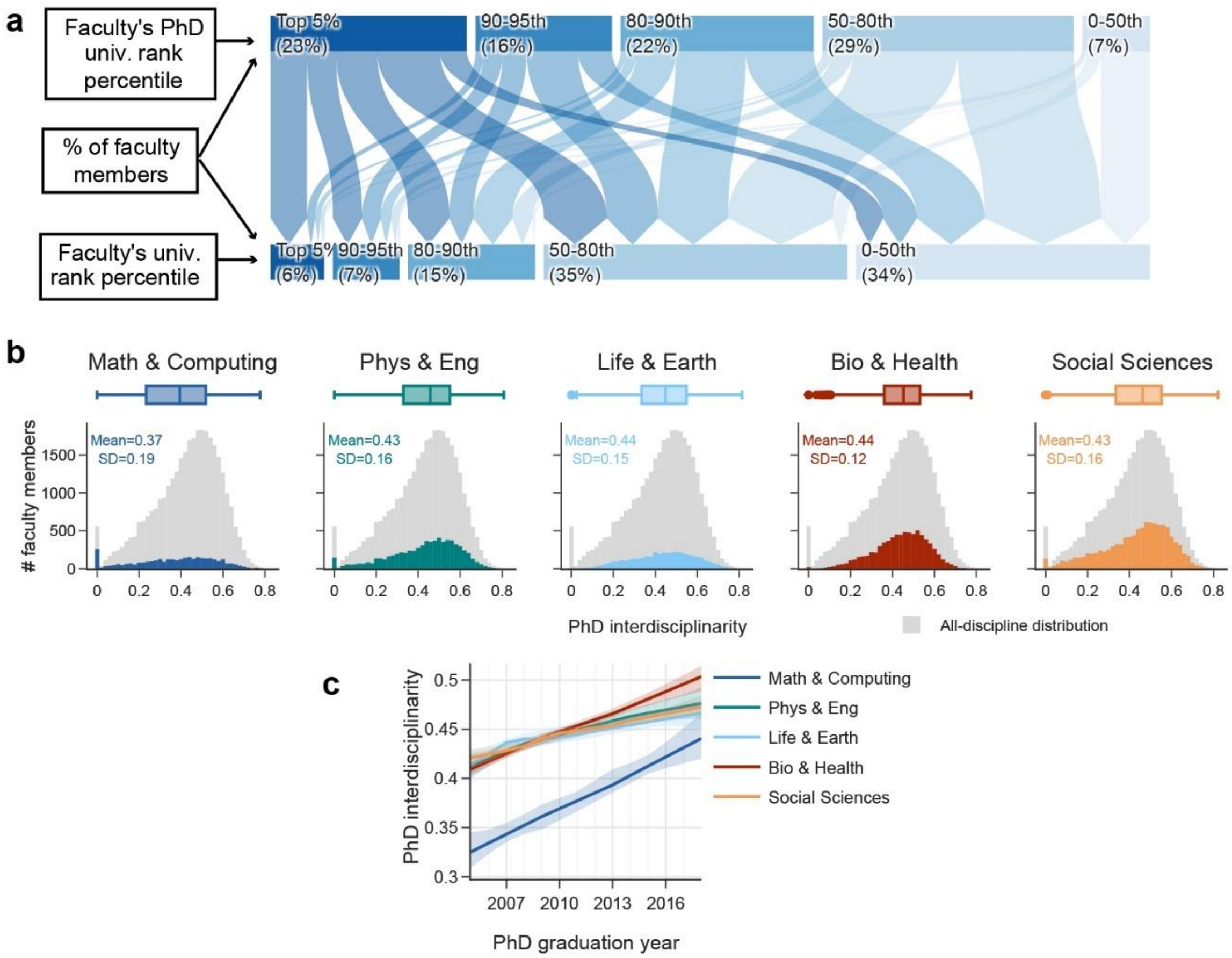

**Fig. S2. Share of faculty newly hired at top 5%~20% universities by Ph.D. graduation year.** Trend lines were generated using LOWESS smoothing, with shaded areas representing 95% bootstrap confidence intervals.

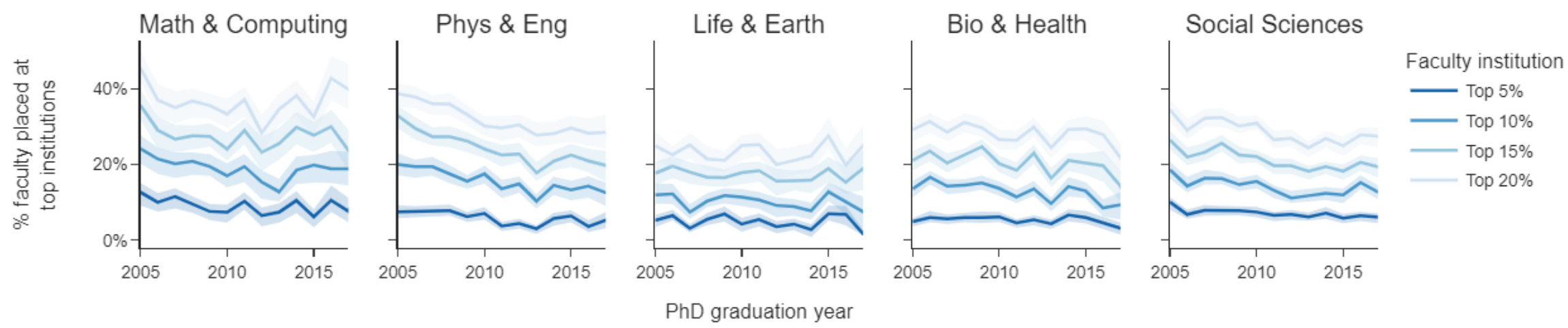

**Fig. S3. New faculty share at the top 5%~20% of universities over Ph.D. interdisciplinarity across fields.** Shaded areas show 95% bootstrap confidence intervals.

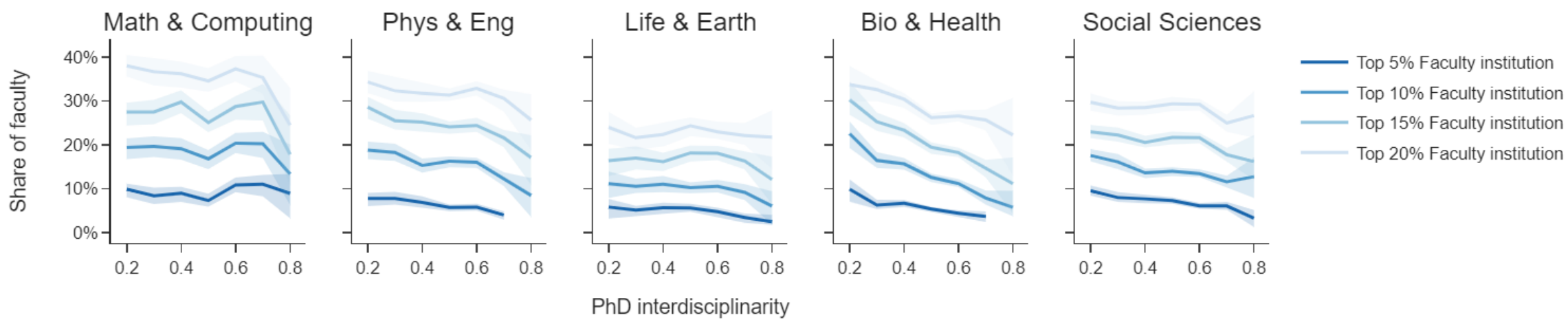

**Fig. S4. Yearly average paper-level interdisciplinarity of faculty by faculty placement across fields.** Shaded areas show 95% confidence intervals. **a-d.** top university rank thresholds of 5%, 10%,15%, and 20%.

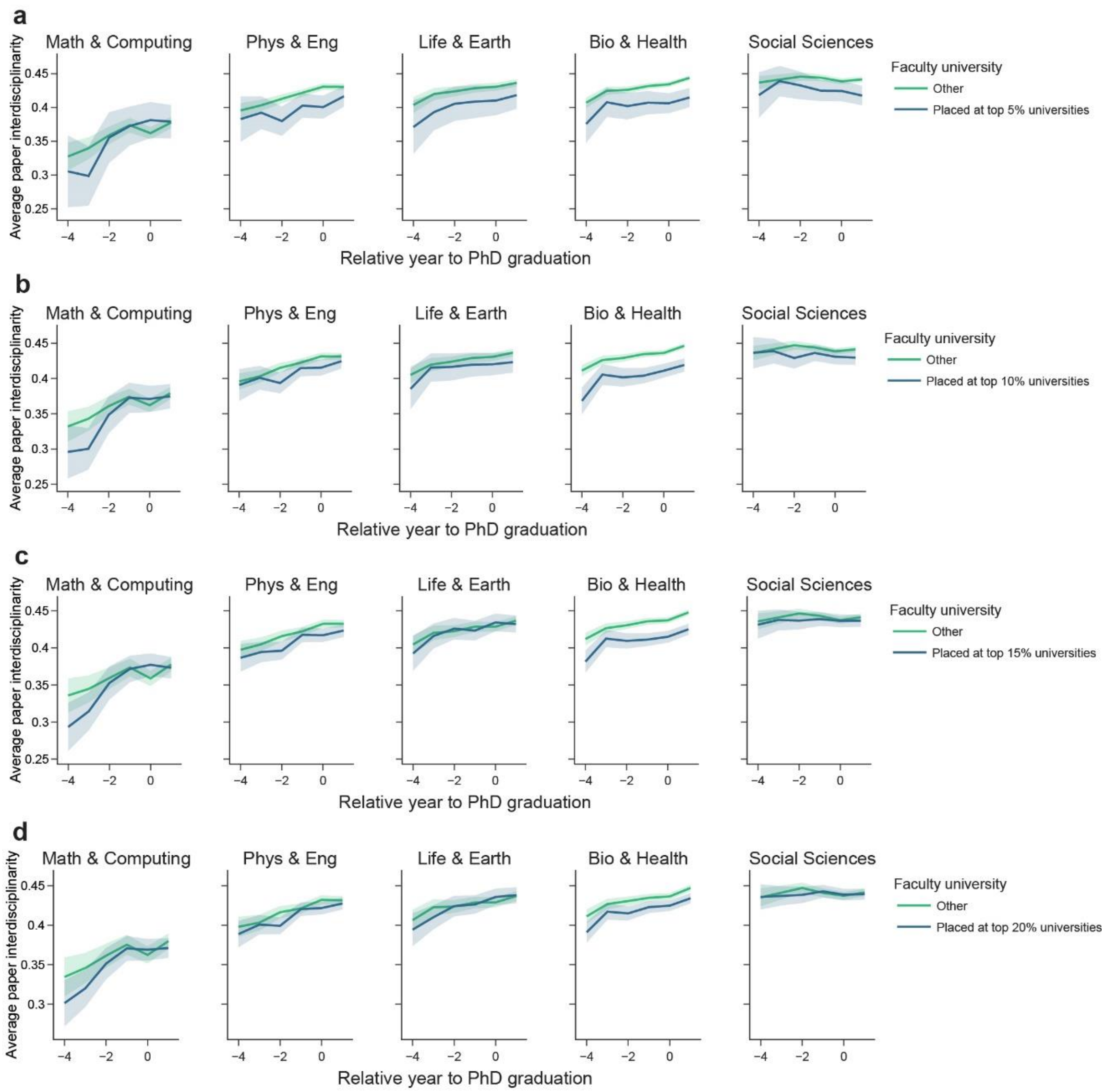

**Fig. S5. Standardized Ph.D. interdisciplinarity. a.** Odds ratios of standardized Ph.D. interdisciplinarity for faculty placement at top-ranked universities. We estimated a series of Logistic regression specifications with various combinations of top university definitions (as top 5%, 10%, 15%, and 20% in respective subfields) and controls (Set 1: Ph.D. university rank and Ph.D. graduation year; Set 2: Set 1+personal and performance features; Set 3: Set 2+advisor features; see **Data and methods**). *** $p < 0.001$, ** $p < 0.01$, * $p < 0.05$. **b-c.** Probability changes of placement types over interdisciplinarity using top university rank thresholds of 5%, 15%, and 20%. Probability changes were estimated by the marginal effects from multinomial logistic regression with a full set of control variables.

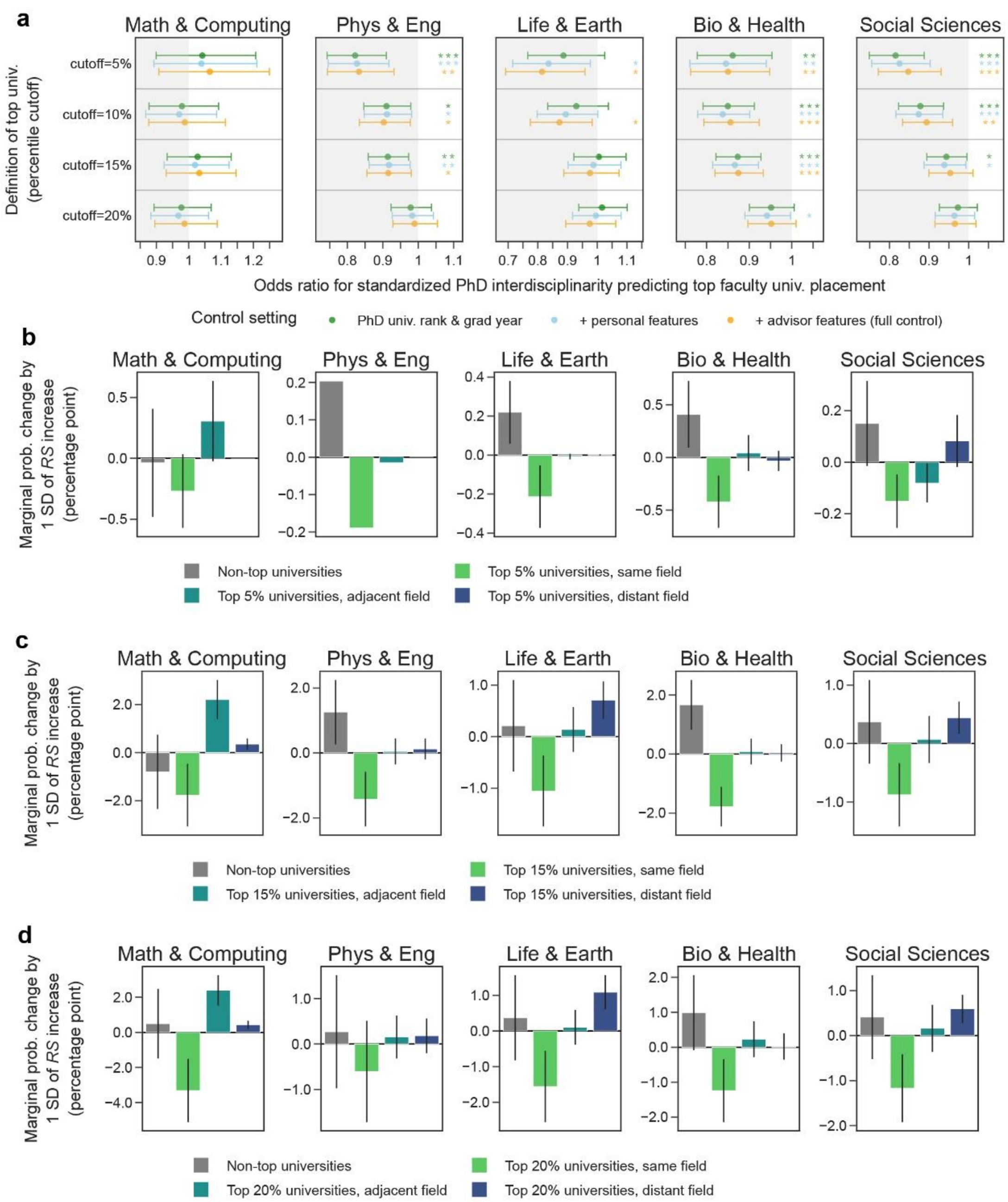

**Fig. S6. Raw Ph.D. interdisciplinarity. a.** Odds ratios of raw Ph.D. interdisciplinarity for faculty placement at top universities by fields. See **Fig. S5a** for chart configuration. *** $p < 0.001$, ** $p < 0.01$, * $p < 0.05$. **b.** Probability changes of placement types over 0.1 Rao-Stirling Index (*RS*) of Ph.D. interdisciplinarity. Probability changes were estimated by the marginal effects from multinomial logistic regression with a full set of control variables. Error bars represent 95% confidence intervals.

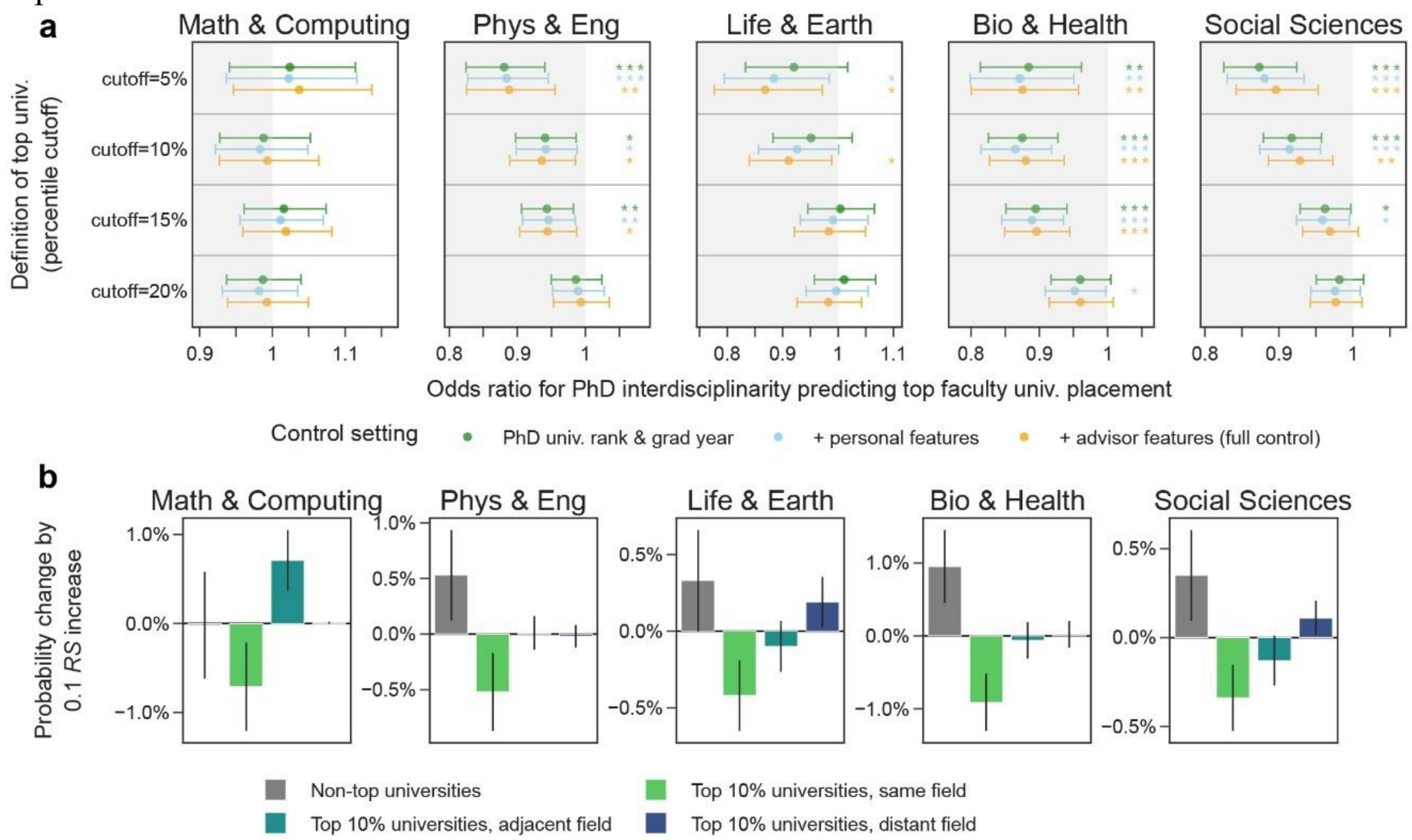

**Fig. S7. Association between Ph.D. interdisciplinarity and faculty university rank across the 10th to 99th quantiles, analyzed by field.** The figure summarizes results from 450 quantile regression models, with estimates and 95% confidence intervals adjusted for all control variables. Red segments represent statistically significant negative associations ($p < 0.05$), indicating that higher Ph.D. interdisciplinarity corresponds to lower faculty university ranks. Gray segments represent non-significant results ($p \geq 0.05$).

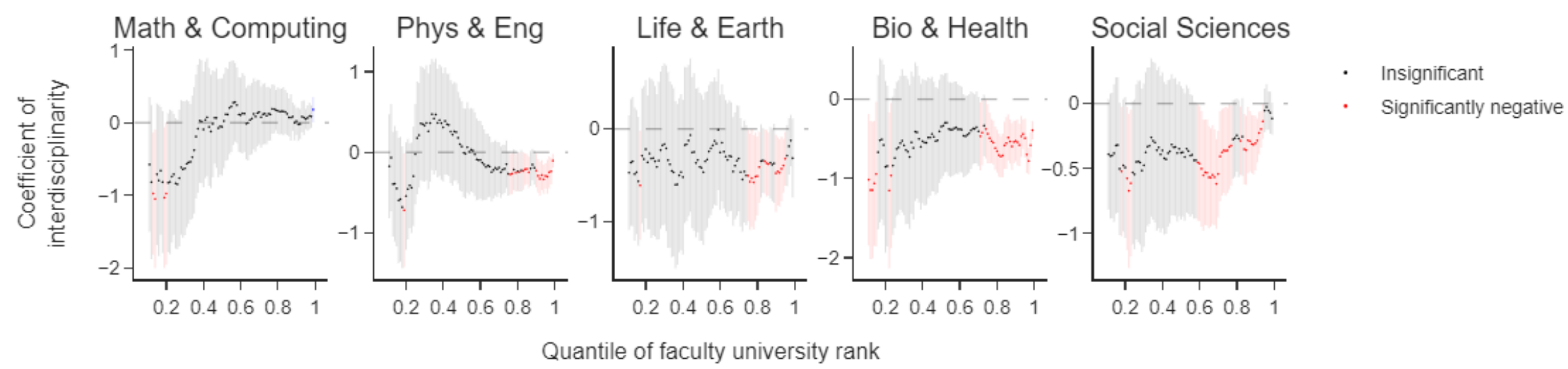

**Fig. S8. Marginal effects of Ph.D. interdisciplinarity on the predicted probability of being placed at top universities at various levels of Ph.D. university ranks.** Ph.D. university ranks are within individuals' Ph.D. subfields and range from the 0th to 100th percentile. We added an interaction term between Ph.D. interdisciplinarity and the individual's Ph.D. university ranking percentile in their respective subfields to the logistic regression model in **Fig. 1b** with a full set of controls. Then, we estimated the average marginal effects of Ph.D. interdisciplinarity on the predicted probability of being placed at top universities at various levels of Ph.D. university ranks.

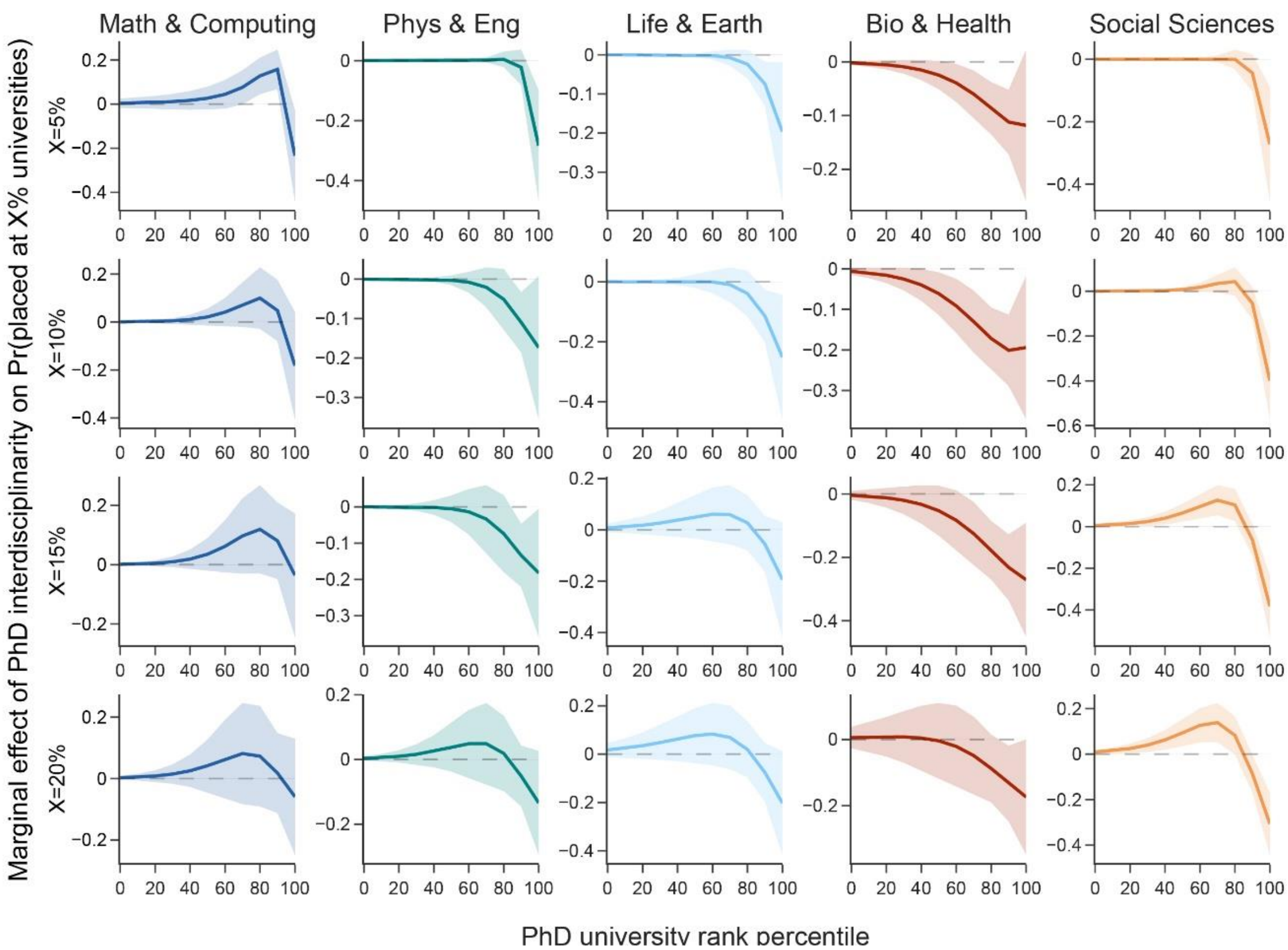

**Fig. S9. Placement types. a.** Average Ph.D. interdisciplinarity by placement types across fields. **b.** Proportion of faculty by different field placement types over Ph.D. interdisciplinarity across fields.

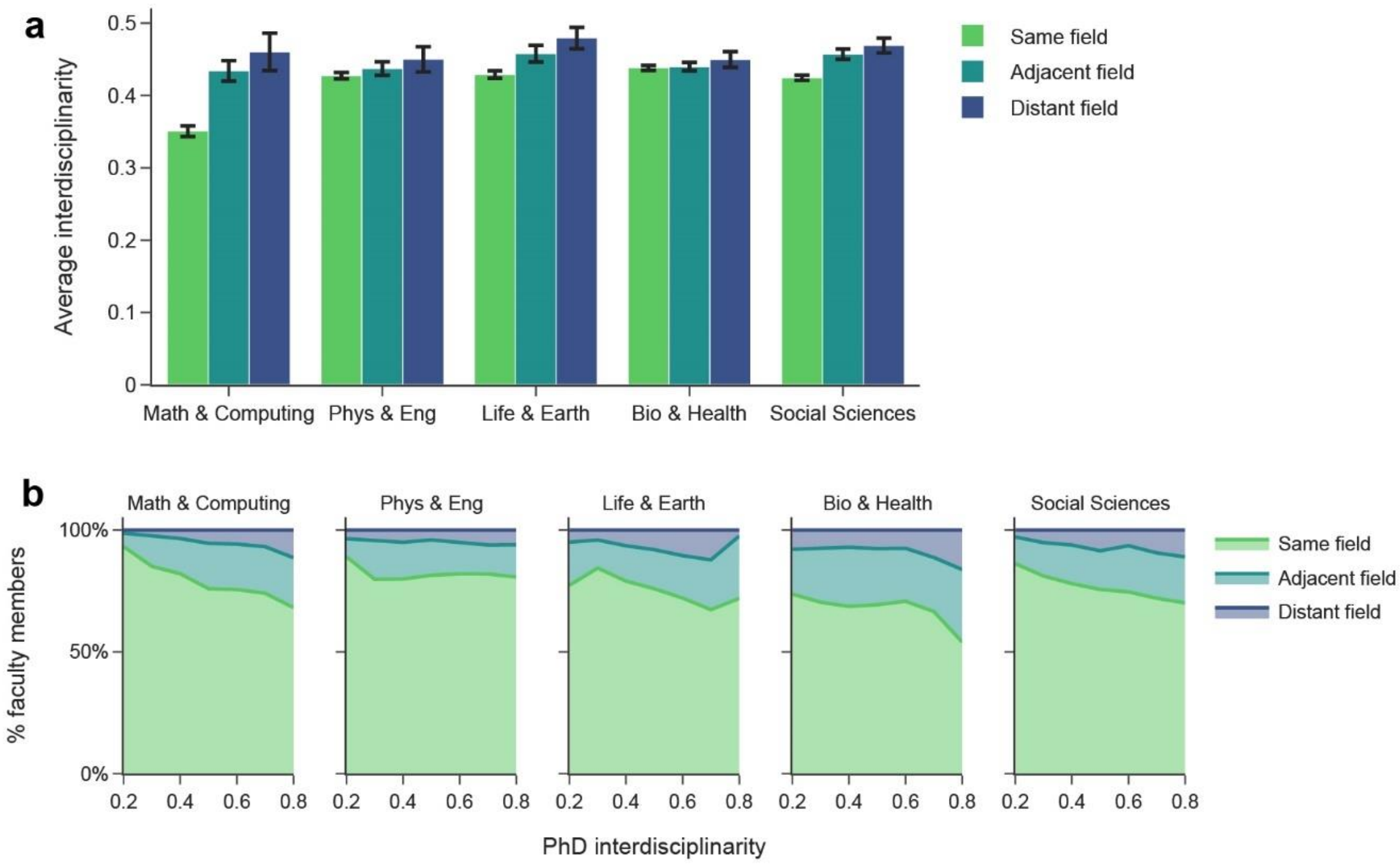

**Fig. S10. Deviation of newly hired faculty's Ph.D. research from existing faculty's research in the same subfield, by placement university rank and Ph.D. interdisciplinarity**. The Ph.D. interdisciplinarity of newly hired faculty is divided into seven intervals (~0.2, 0.3-0.4, …, 0.7~0.8, 0.8~, due to sparse data at both ends), and placement university rank is divided into ten equal intervals. The color of each cell indicates the average deviation for the corresponding faculty and university groups.

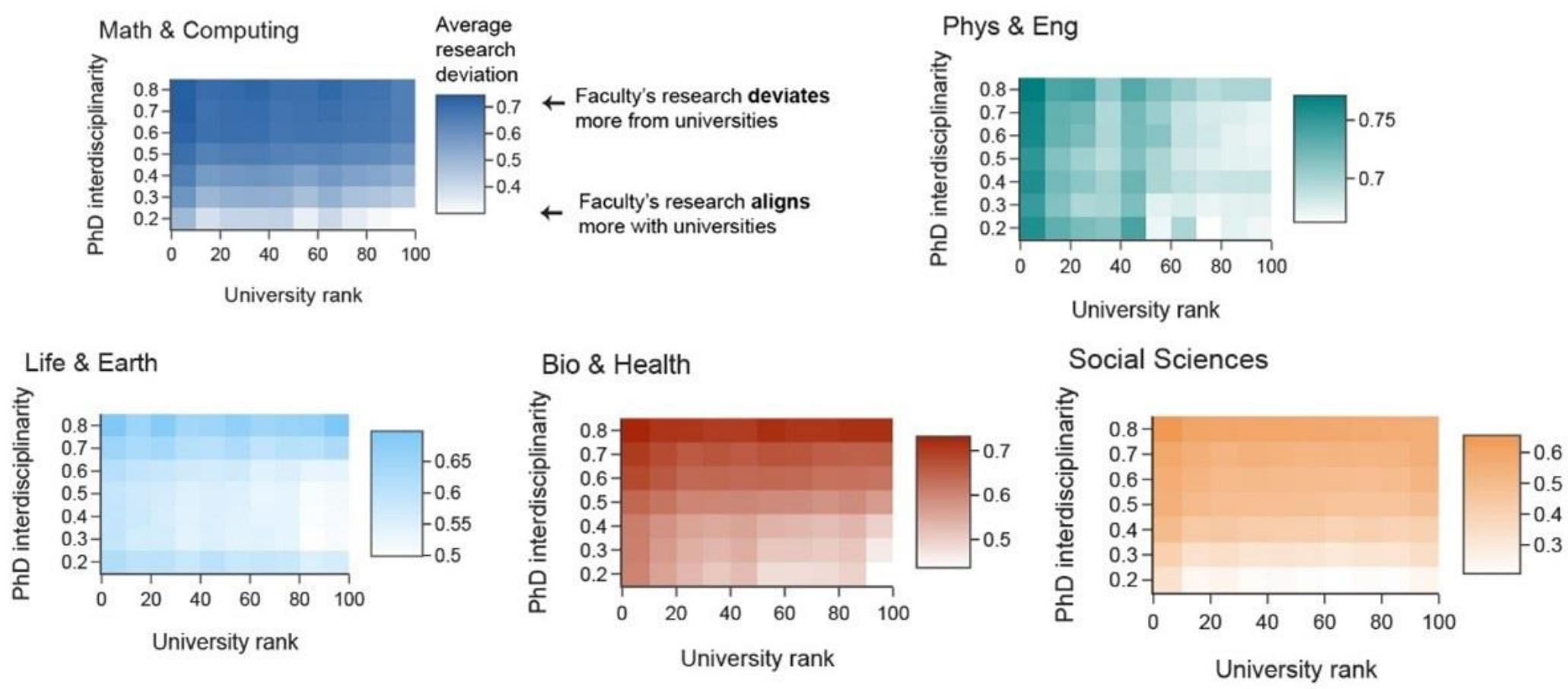

**Fig. S11. Cumulative share of men and women faculty at top X% faculty universities.**

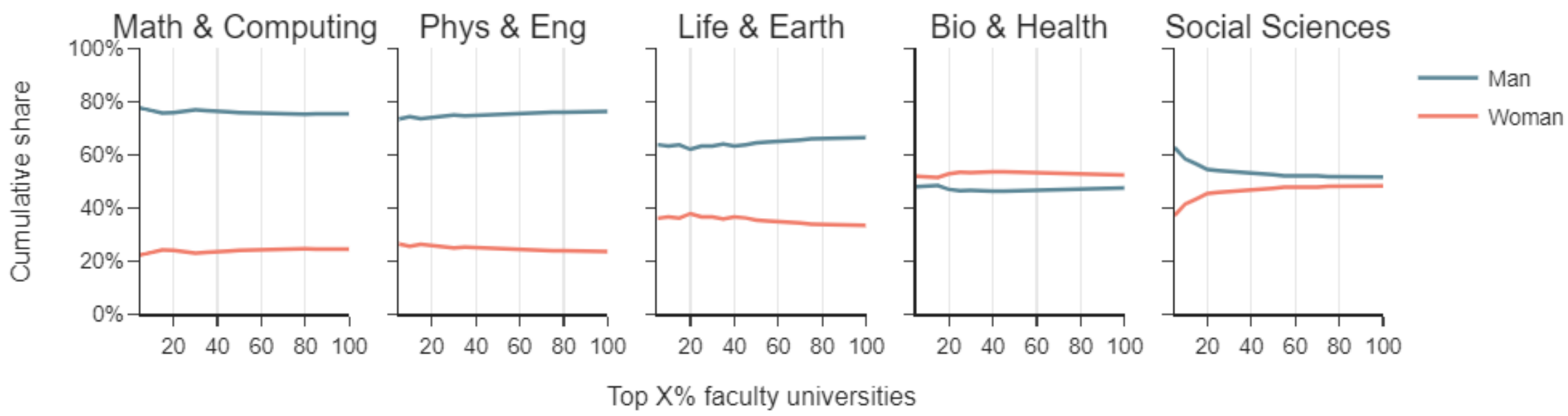

**Fig. S12. Annual productivity during Ph.D. training and 10 years after graduation, categorized by field and Ph.D. interdisciplinarity level. a.** At non-top-10% universities. Ph.D. interdisciplinarity level categorized by above or below mean values. **b.** Ph.D. interdisciplinarity level categorized by equal-size quintiles. The five quintiles were divided within each subfield and Ph.D. graduation year cohort, where quintile 5 denotes the highest level of Ph.D. interdisciplinarity. Trend lines were obtained using LOWESS smoothing. Shaded areas show 95% bootstrap confidence intervals.

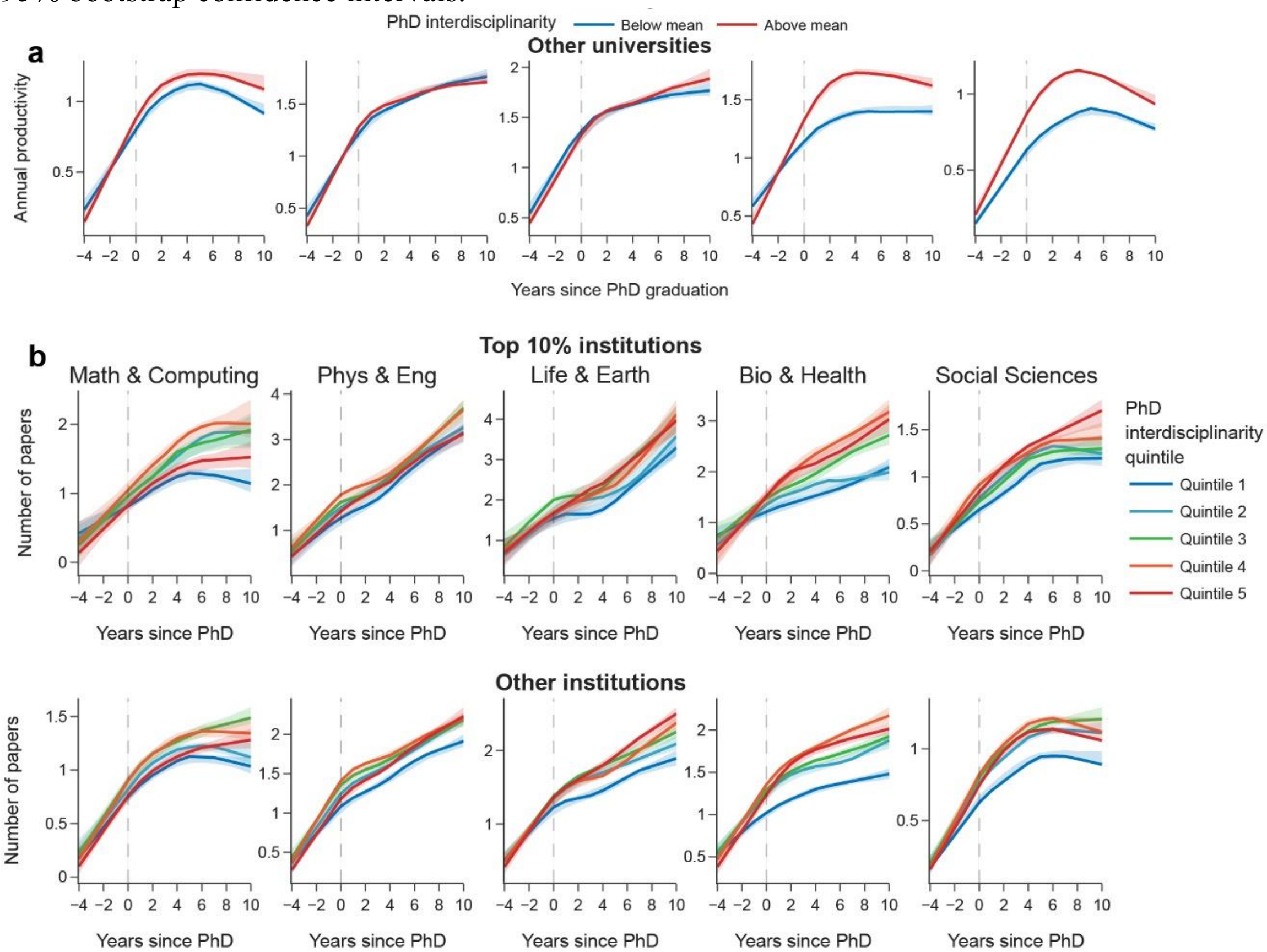

**Fig. S13. Changes in productivity and normalized citation per 1 standard deviation (SD) increase in Ph.D. interdisciplinarity by field and faculty university rank.** Changes were estimated by the marginal effects from regressions. Shaded areas show 95% bootstrap confidence intervals. **a-c.** Changes in the number of papers with thresholds of top faculty universities changed to 5%, 15%, and 20%. **d.** Changes in normalized citation counts. Citations are normalized by the average citation counts within each discipline and publication year.

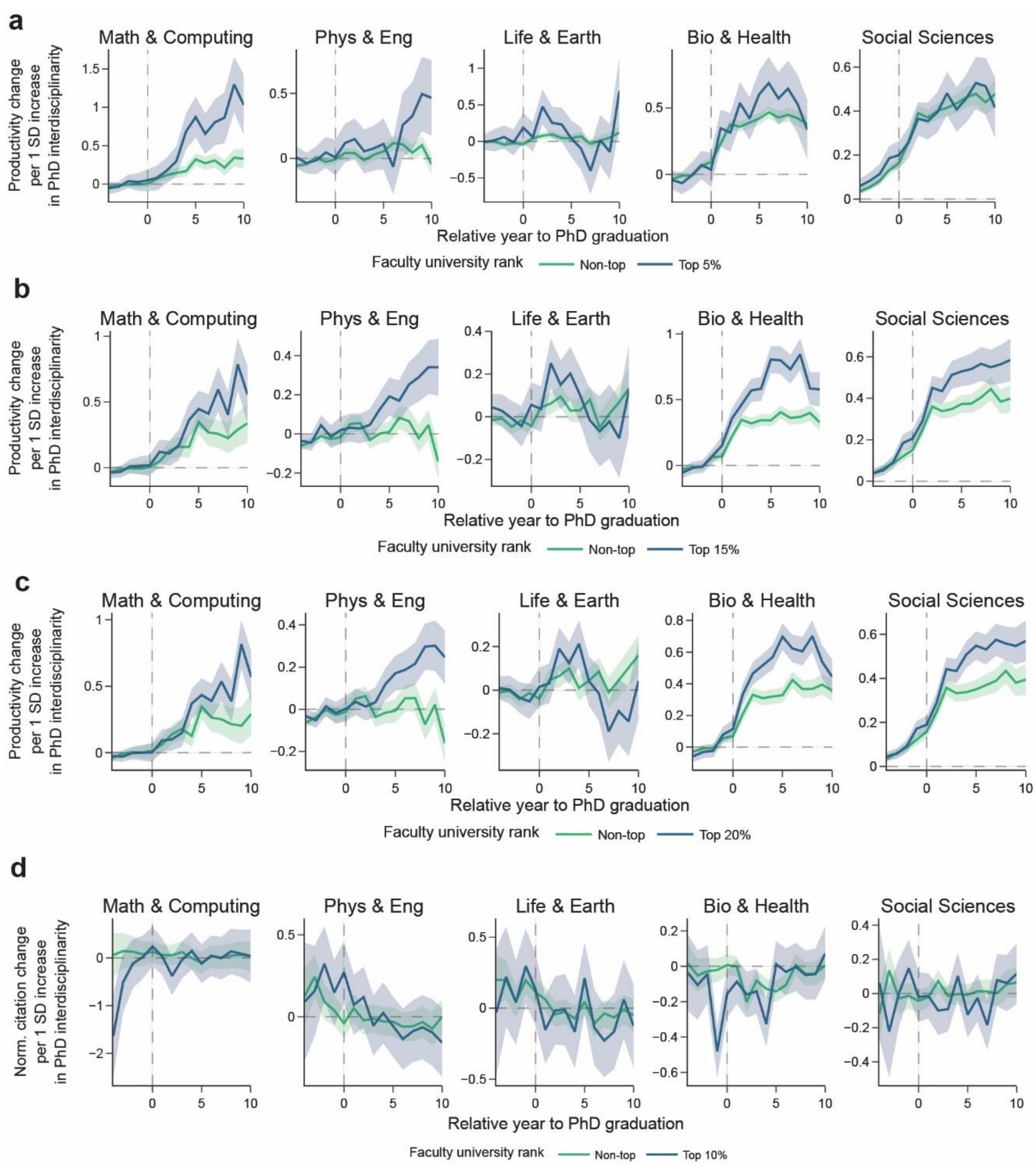

**Fig. S14. Among all who were faculty X years after their Ph.D. graduation, the proportion of faculty remaining affiliated with their initial placement university.** Estimates are based on newly hired faculty who earned their Ph.D. after 2010, ensuring their initial faculty placement is trackable using AARC data (2011–2020).

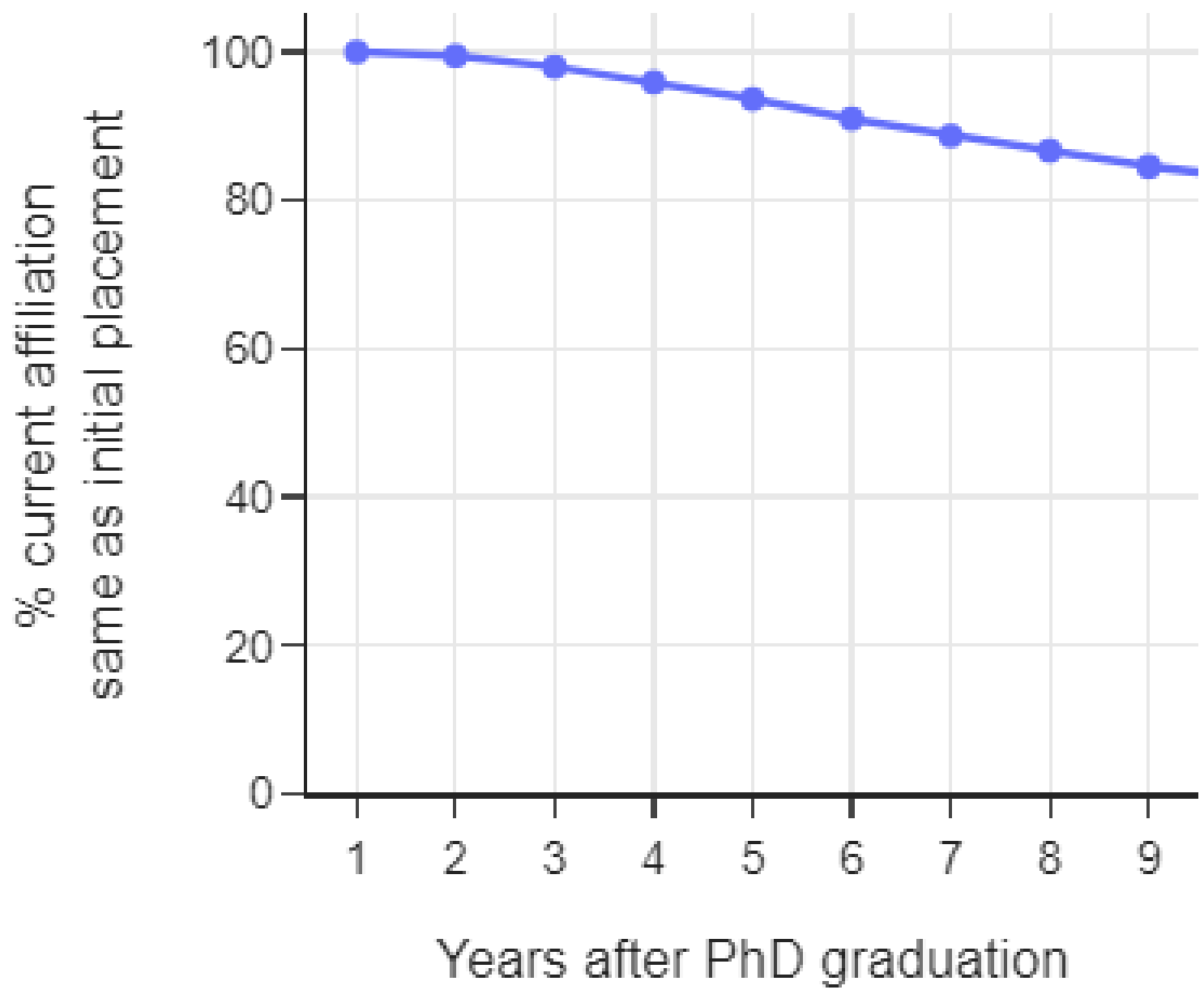

**Fig. S15. Test on faculty members who graduated after 2009 and had the earliest faculty records no more than two years after their Ph.D. graduation. a.** Odds ratios of Ph.D. interdisciplinarity for faculty placement at top universities by fields. See **Fig. S5e** for chart configuration. *** $p < 0.001$, ** $p < 0.01$, * $p < 0.05$. **b.** Probability changes of placement types over Ph.D. interdisciplinarity. Probability changes were estimated by the marginal effects from multinomial logistic regression with a full set of control variables. Error bars represent 95% confidence intervals.

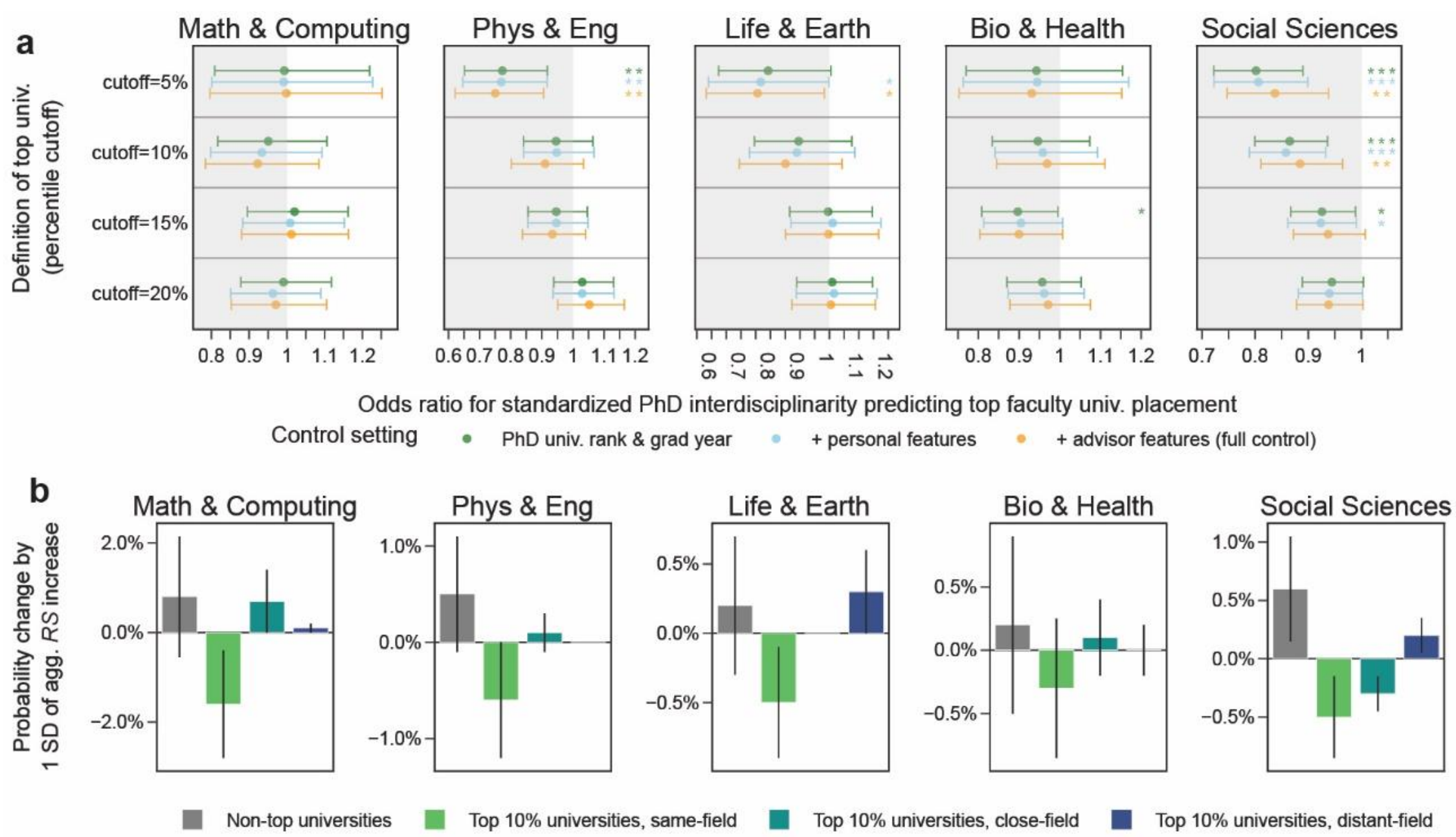

**Fig. S16. Pooled Ph.D. interdisciplinarity. a.** Odds ratios of pooled Ph.D. interdisciplinarity for faculty placement at top universities by fields. See **Fig. S5e** for chart configuration. *** $p < 0.001$, ** $p < 0.01$, * $p < 0.05$. **b.** Probability changes of placement types over pooled Ph.D. interdisciplinarity. Probability changes were estimated by the marginal effects from multinomial logistic regression with a full set of control variables. Error bars represent 95% confidence intervals.

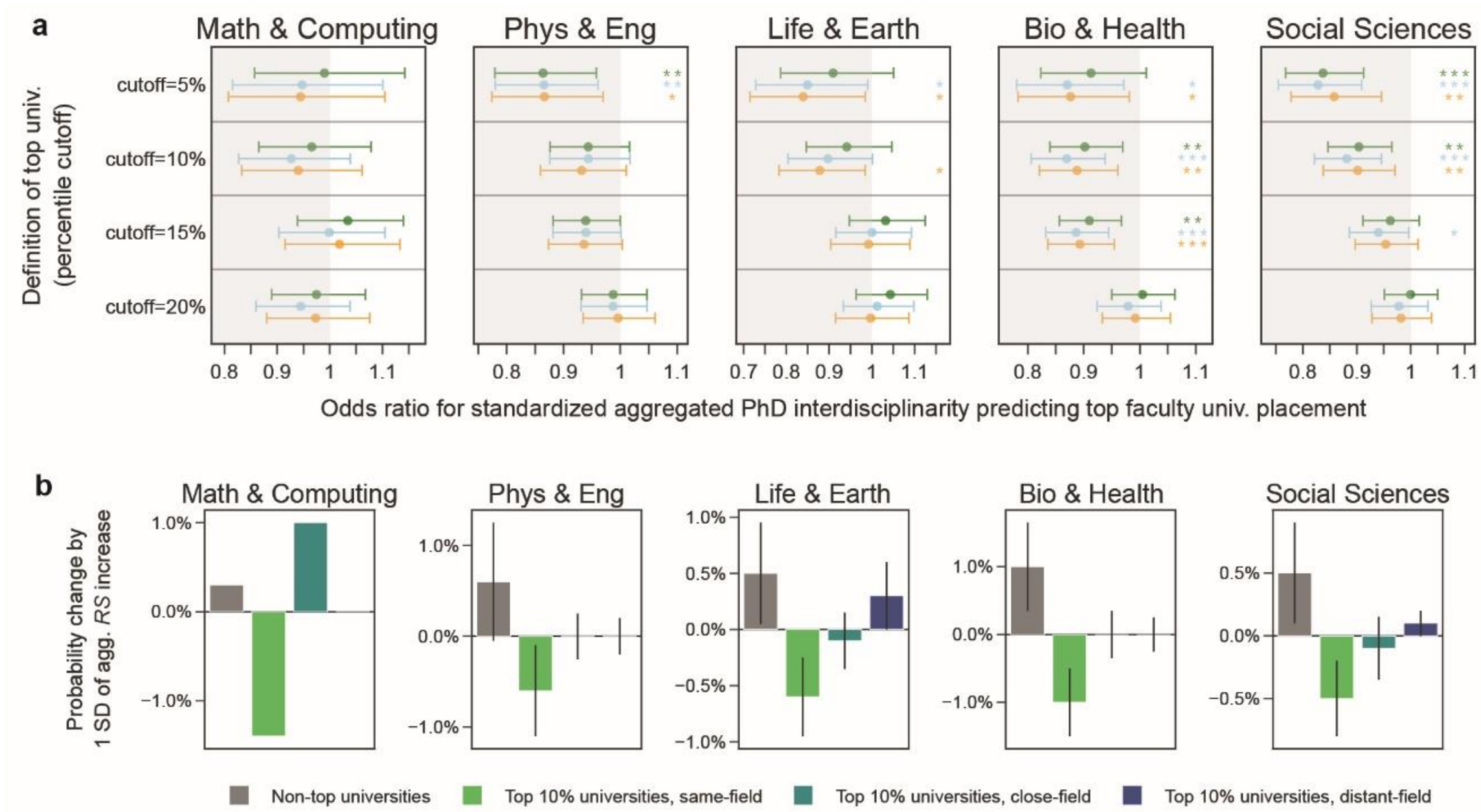

**Fig. S17. Top universities in the world university rankings.** World university rankings include the 2020 Academic Ranking of World Universities (ARWU), QS World University Rankings (QS), Times Higher Education World University Rankings (Times), and U.S. News & World Report Best Global Universities Rankings (US News). **a.** Odds ratios of standardized Ph.D. interdisciplinarity for faculty placement at top universities defined by four world university rankings by fields. Here, top universities are U.S. universities ranked within the top 20 nationwide in one of the world university rankings in 2020. *** $p<0.001$; ** $p<0.01$; * $p<0.05$. **b.** Probability changes of placement types (top 20 universities in one of the world university rankings vs others) over Ph.D. interdisciplinarity.

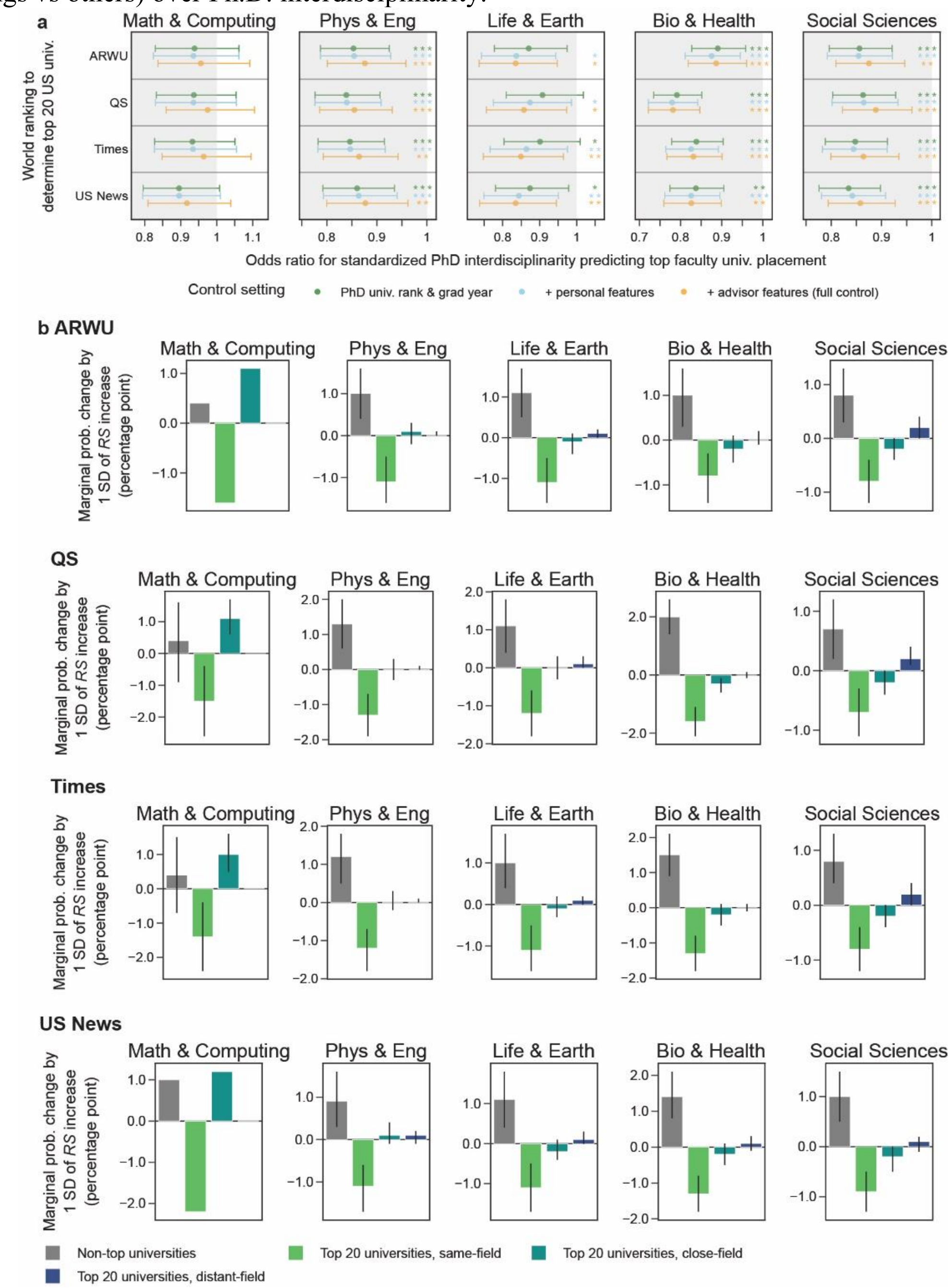

**Fig. S18. Subfield distance. a.** Ph.D.-to-faculty field mobility network. Edge weights are normalized numbers of faculty members who moved from a source subfield of Ph.D. study to a target subfield of the faculty department. Subfields are colored by their affiliated fields. **b.** Cumulative distribution of faculty members who moved their subfields from Ph.D. to faculty over subfield distance.

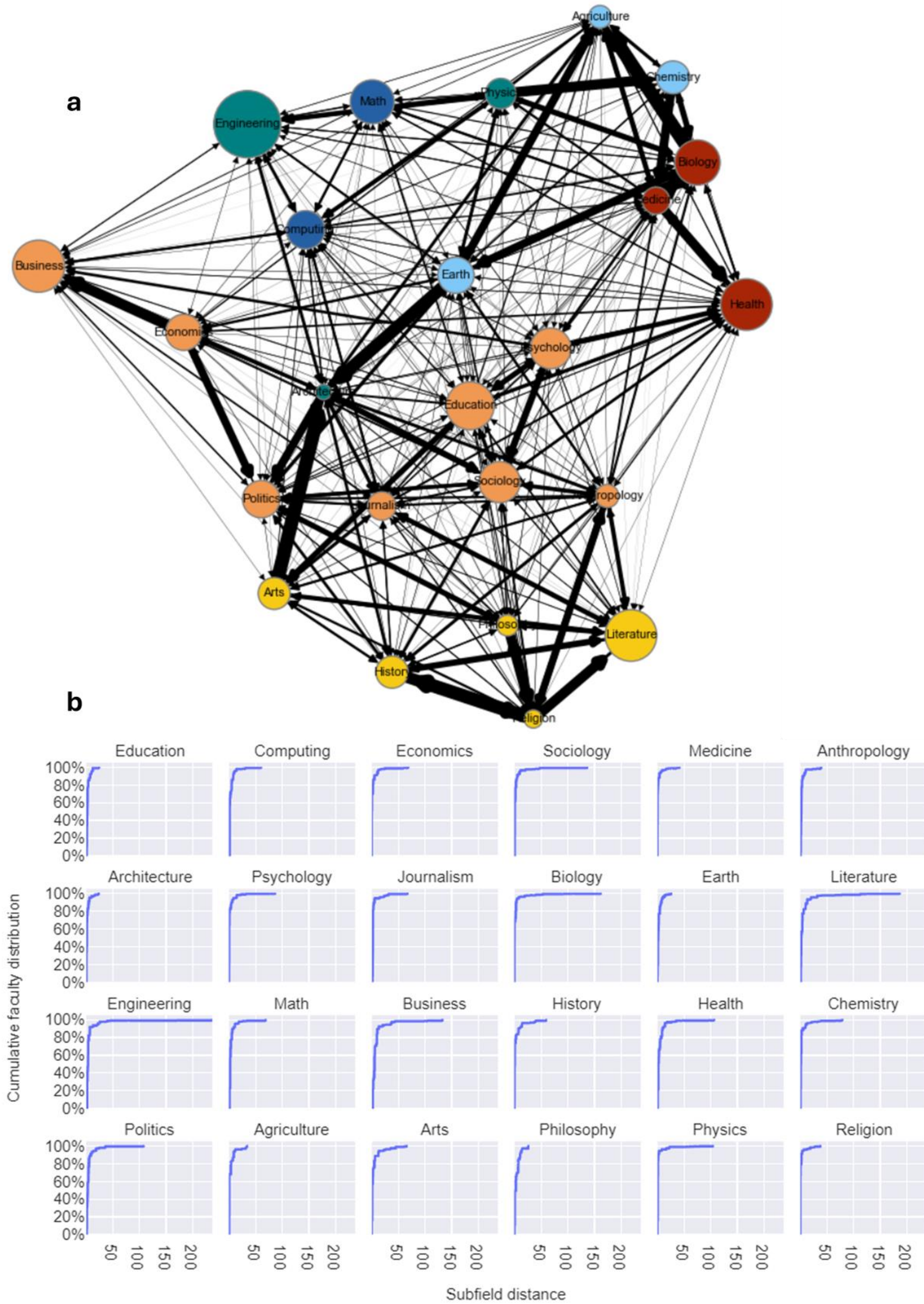

**Table S1**. **Distributions of new faculty members and descriptive statistics of the analytical sample.**

| Distribution | | |
|---|---|---|
| | **Number of faculty** | **%** |
| **PhD field (*subfield*)** | | |
| Math & Computing | 3306 | 10.4 |
| • *Computing* | *1383* | *4.4* |
| • *Math* | *1923* | *6.1* |
| Phys & Eng | 6741 | 21.2 |
| • *Architecture* | *178* | *0.6* |
| • *Engineering* | *5166* | *16.3* |
| • *Physics* | *1397* | *4.4* |
| Life & Earth | 4042 | 12.7 |
| • *Agriculture* | *612* | *1.9* |
| • *Chemistry* | *1540* | *4.8* |
| • *Earth* | *1890* | *5.9* |
| Bio & Health | 7272 | 22.9 |
| • *Biology* | *3345* | *10.5* |
| • *Health* | *2526* | *7.9* |
| • *Medicine* | *1401* | *4.4* |
| Social Sciences | 10414 | 32.8 |
| • *Anthropology* | *423* | *1.3* |
| • *Business* | *1737* | *5.5* |
| • *Economics* | *1009* | *3.2* |
| • *Education* | *1145* | *3.6* |
| • *Journalism* | *695* | *2.2* |
| • *Politics* | *1100* | *3.5* |
| • *Psychology* | *2524* | *7.9* |
| • *Sociology* | *1781* | *5.6* |
| **Faculty field (*subfield*)** | | |
| Math & Computing | 3215 | 10.1 |
| • *Computing* | *1441* | *4.5* |
| • *Math* | *1774* | *5.6* |
| Phys & Eng | 6387 | 20.1 |
| • *Architecture* | *196* | *0.6* |
| • *Engineering* | *5076* | *16* |
| • *Physics* | *1115* | *3.5* |
| Life & Earth | 4092 | 12.9 |
| • *Agriculture* | *828* | *2.6* |
| • *Chemistry* | *1366* | *4.3* |
| • *Earth* | *1898* | *6* |
| Bio & Health | 7814 | 24.6 |
| • *Biology* | *3154* | *9.9* |
| • *Health* | *3154* | *9.9* |
| • *Medicine* | *1506* | *4.7* |
| Social Sciences | 10117 | 31.8 |
| • *Anthropology* | *359* | *1.1* |
| • *Business* | *2203* | *6.9* |
| • *Economics* | *679* | *2.1* |
| • *Education* | *1124* | *3.5* |
| • *Journalism* | *643* | *2* |
| • *Politics* | *1231* | *3.9* |
| • *Psychology* | *1995* | *6.3* |
| • *Sociology* | *1883* | *5.9* |
| Arts & Humanities | 150 | 0.5 |
| • *Arts* | *29* | *0.1* |
| • *History* | *14* | *0* |
| • *Literature* | *79* | *0.2* |
| • *Philosophy* | *20* | *0.1* |
| • *Religion* | *8* | *0* |
| **Ph.D. graduation year** | | |
| 2005 | 2452 | 7.7 |
| 2006 | 2635 | 8.3 |
| 2007 | 2856 | 9 |
| 2008 | 3000 | 9.4 |
| 2009 | 3088 | 9.7 |
| 2010 | 2742 | 8.6 |
| 2011 | 2657 | 8.4 |
| 2012 | 2214 | 7 |
| 2013 | 2414 | 7.6 |
| 2014 | 2269 | 7.1 |
| 2015 | 2087 | 6.6 |
| 2016 | 1623 | 5.1 |
| 2017 | 1219 | 3.8 |
| 2018 | 519 | 1.6 |
| **Gender** | | |
| Woman | 12585 | 39.6 |
| Man | 19190 | 60.4 |
| **Ph.D. advisor gender** | | |
| Woman | 7649 | 26.3 |
| Man | 21447 | 73.7 |
| **Descriptive statistics** | | |
| | **Mean** | **Standard deviation** |
| Ph.D. interdisciplinarity (Rao-Stirling index) | 0.43 | 0.16 |
| Faculty university rank percentile | 60.16 | 26.49 |
| Ph.D. university rank percentile | 80.56 | 18.11 |
| Ph.D. average normalized citations | 1.26 | 2.04 |
| Ph.D. publications | 4.98 | 4.42 |
| Ph.D. collaborator number | 7.06 | 7.05 |
| Advisor past-five-year advisor publications before Ph.D. start year | 9.61 | 15.55 |
| Advisor seniority before Ph.D. start year | 15.56 | 9.65 |

**Table S2. Logistic regression of placement types (reference group: same-field stayers) for faculty placement at top universities by fields.** *** $p<0.001$; ** $p<0.01$; * $p<0.05$.

| | **Math & Computing** | **Phys & Eng** | **Life & Earth** | **Bio & Health** | **Social Sciences** |
|---|---|---|---|---|---|
| ***Top university threshold = 95th*** | | | | | |
| Adjacent field | 0.473* (0.194) | 0.584*** (0.151) | 0.076 (0.241) | -0.102 (0.139) | 0.507*** (0.120) |
| Distant field | 0.748* (0.364) | -0.051 (0.312) | 0.535* (0.237) | 0.871*** (0.158) | 0.575*** (0.162) |
| ***Top university threshold = 90th*** | | | | | |
| Adjacent field | 0.424** (0.153) | 0.349** (0.107) | 0.088 (0.168) | -0.037 (0.094) | 0.460*** (0.094) |
| Distant field | 0.521 (0.302) | 0.214 (0.198) | 0.671*** (0.187) | 0.497*** (0.129) | 0.495*** (0.126) |
| ***Top university threshold = 85th*** | | | | | |
| Adjacent field | 0.430** (0.139) | 0.306** (0.096) | 0.116 (0.134) | -0.043 (0.080) | 0.385*** (0.081) |
| Distant field | 0.689* (0.270) | 0.603*** (0.168) | 0.792*** (0.161) | 0.487*** (0.115) | 0.508*** (0.110) |
| ***Top university threshold = 80th*** | | | | | |
| Adjacent field | 0.214 (0.131) | 0.196* (0.091) | -0.012 (0.122) | -0.077 (0.071) | 0.495*** (0.076) |
| Distant field | 0.505 (0.261) | 0.558*** (0.160) | 0.849*** (0.154) | 0.461*** (0.106) | 0.442*** (0.103) |
| Full controls | Yes | Yes | Yes | Yes | Yes |

**Table S3. Logistic regression of interdisciplinarity for faculty placement at top 10% universities by fields and subgroups of placement type.** *** p<0.001; ** p<0.01; * p<0.05.

| **Field** | **Placement type** | **Coefficient** | **SE** | **p** | **CI lower** | **CI upper** |
|---|---|---|---|---|---|---|
| Math & Computing | Same field | -0.593 | 0.398 | 0.136 | -1.373 | 0.186 |
| | Adjacent field | 1.326 | 1.175 | 0.259 | -0.976 | 3.629 |
| | Distant field | -3.245 | 2.176 | 0.136 | -7.511 | 1.020 |
| Phys & Eng | Same field | -0.652* | 0.294 | 0.026 | -1.227 | -0.076 |
| | Adjacent field | -0.703 | 0.742 | 0.343 | -2.158 | 0.751 |
| | Distant field | -0.705 | 1.453 | 0.628 | -3.554 | 2.143 |
| Life & Earth | Same field | -1.201* | 0.505 | 0.017 | -2.191 | -0.211 |
| | Adjacent field | -1.778 | 0.970 | 0.067 | -3.679 | 0.123 |
| | Distant field | 1.685 | 1.416 | 0.234 | -1.091 | 4.460 |
| Bio & Health | Same field | -1.485*** | 0.389 | 0.000 | -2.247 | -0.722 |
| | Adjacent field | -0.774 | 0.666 | 0.245 | -2.079 | 0.530 |
| | Distant field | -1.465 | 0.905 | 0.105 | -3.238 | 0.308 |
| Social Sciences | Same field | -0.826*** | 0.279 | 0.003 | -1.374 | -0.279 |
| | Adjacent field | -1.469* | 0.584 | 0.012 | -2.614 | -0.325 |
| | Distant field | -0.440 | 1.135 | 0.698 | -2.664 | 1.784 |

**Table S4. Number of universities and top universities in the rankings by subfield.**

| Field | Subfield | # universities in the ranking | # top 5% universities | # top 10% universities |
|---|---|---|---|---|
| Math & Computing | Computing | 248 | 13 | 25 |
| | Math | 247 | 13 | 25 |
| Phys & Eng | Architecture | 110 | 6 | 11 |
| | Engineering | 232 | 12 | 24 |
| | Physics | 221 | 12 | 23 |
| Life & Earth | Agriculture | 81 | 5 | 9 |
| | Chemistry | 249 | 13 | 25 |
| | Earth | 216 | 11 | 22 |
| Bio & Health | Biology | 286 | 15 | 29 |
| | Health | 248 | 13 | 25 |
| | Medicine | 167 | 9 | 17 |
| Social Sciences | Anthropology | 163 | 9 | 17 |
| | Business | 219 | 11 | 22 |
| | Economics | 207 | 11 | 21 |
| | Education | 228 | 12 | 23 |
| | Journalism | 169 | 9 | 17 |
| | Politics | 214 | 11 | 22 |
| | Psychology | 278 | 14 | 28 |
| | Sociology | 228 | 12 | 23 |
| Humanities | Arts | 205 | 11 | 21 |
| | History | 217 | 11 | 22 |
| | Literature | 226 | 12 | 23 |
| | Philosophy | 181 | 10 | 19 |
| | Religion | 121 | 7 | 13 |

**Table S5. Covariate balance checks for propensity score matching models.**

| Variable | Match | Women Mean | Men Mean | % Bias | % Bias Reduction | t-value | p-value |
|---|---|---|---|---|---|---|---|
| **Math & Computing** | | | | | | | |
| PhD univ. ranking | Unmatched | 83.28 | 84.82 | -9.7 | — | -2 | 0.046 |
| | Matched | 83.28 | 84.31 | -6.4 | 33.5 | -1.05 | 0.294 |
| PhD avg. citations | Unmatched | 1.356 | 1.567 | -8 | — | -1.44 | 0.15 |
| | Matched | 1.356 | 1.375 | -0.7 | 90.9 | -0.18 | 0.859 |
| PhD num. papers | Unmatched | 4.593 | 5.145 | -15.7 | — | -3.12 | 0.002 |
| | Matched | 4.593 | 4.703 | -3.1 | 80 | -0.55 | 0.582 |
| PhD num. collaborators | Unmatched | 7.056 | 7.111 | -0.9 | — | -0.18 | 0.855 |
| | Matched | 7.056 | 7.004 | 0.8 | 7.8 | 0.14 | 0.889 |
| Woman advisor | Unmatched | 0.235 | 0.157 | 19.7 | — | 4.14 | 0 |
| | Matched | 0.235 | 0.171 | 16.1 | 18.3 | 2.6 | 0.009 |
| Seniority Q2 | Unmatched | 0.226 | 0.228 | -0.7 | — | -0.14 | 0.888 |
| | Matched | 0.226 | 0.226 | -0.1 | 86.9 | -0.02 | 0.988 |
| Seniority Q3 | Unmatched | 0.237 | 0.228 | 1.9 | — | 0.39 | 0.695 |
| | Matched | 0.237 | 0.229 | 1.8 | 5.6 | 0.3 | 0.764 |
| Seniority Q4 | Unmatched | 0.168 | 0.168 | 0.1 | — | 0.02 | 0.981 |
| | Matched | 0.168 | 0.171 | -0.8 | -588.3 | -0.14 | 0.892 |
| Seniority Q5 | Unmatched | 0.152 | 0.149 | 0.6 | — | 0.13 | 0.898 |
| | Matched | 0.152 | 0.154 | -0.7 | -17 | -0.12 | 0.904 |
| Num. Papers of Advisors | Unmatched | 9.089 | 9.443 | -2.7 | — | -0.51 | 0.613 |
| | Matched | 9.089 | 9.295 | -1.6 | 41.8 | -0.27 | 0.786 |
| **With Ph.D. interdisciplinarity** | | | | | | | |
| PhD interdisciplinarity | Unmatched | 0.4332 | 0.3928 | 24.2 | — | 4.76 | 0 |
| | Matched | 0.4332 | 0.4197 | 8.1 | 66.7 | 1.37 | 0.17 |
| PhD univ. ranking | Unmatched | 83.28 | 84.82 | -9.7 | — | -2 | 0.046 |
| | Matched | 83.28 | 83.98 | -4.4 | 54.9 | -0.71 | 0.48 |
| PhD avg. citations | Unmatched | 1.356 | 1.567 | -8 | — | -1.44 | 0.15 |
| | Matched | 1.356 | 1.368 | -0.4 | 94.5 | -0.11 | 0.913 |
| PhD num. papers | Unmatched | 4.593 | 5.145 | -15.7 | — | -3.12 | 0.002 |
| | Matched | 4.593 | 4.72 | -3.6 | 77 | -0.63 | 0.531 |
| PhD num. collaborators | Unmatched | 7.056 | 7.111 | -0.9 | — | -0.18 | 0.855 |
| | Matched | 7.056 | 7.034 | 0.4 | 61.6 | 0.06 | 0.954 |
| Woman advisor | Unmatched | 0.235 | 0.157 | 19.7 | — | 4.14 | 0 |
| | Matched | 0.235 | 0.188 | 11.8 | 40 | 1.88 | 0.061 |
| Seniority Q2 | Unmatched | 0.226 | 0.228 | -0.7 | — | -0.14 | 0.888 |
| | Matched | 0.226 | 0.223 | 0.5 | 27.7 | 0.08 | 0.934 |
| Seniority Q3 | Unmatched | 0.237 | 0.228 | 1.9 | — | 0.39 | 0.695 |
| | Matched | 0.237 | 0.23 | 1.5 | 20 | 0.25 | 0.8 |
| Seniority Q4 | Unmatched | 0.168 | 0.168 | 0.1 | — | 0.02 | 0.981 |
| | Matched | 0.168 | 0.172 | -1 | -724.3 | -0.16 | 0.871 |
| Seniority Q5 | Unmatched | 0.152 | 0.149 | 0.6 | — | 0.13 | 0.898 |
| | Matched | 0.152 | 0.155 | -1 | -52.6 | -0.16 | 0.875 |
| Num. Papers of Advisors | Unmatched | 9.089 | 9.443 | -2.7 | — | -0.51 | 0.613 |
| | Matched | 9.089 | 9.295 | -1.6 | 41.9 | -0.27 | 0.789 |
| **Physics & Engineering** | | | | | | | |
| PhD univ. ranking | Unmatched | 83.455 | 83.878 | -2.5 | — | -0.79 | 0.428 |
| | Matched | 83.442 | 83.663 | -1.3 | 47.9 | -0.33 | 0.74 |
| PhD avg. citations | Unmatched | 1.677 | 1.733 | -2.4 | — | -0.77 | 0.444 |
| | Matched | 1.667 | 1.708 | -1.8 | 25.9 | -0.46 | 0.648 |
| PhD num. papers | Unmatched | 6.122 | 6.965 | -19.7 | — | -6.03 | 0 |
| | Matched | 6.112 | 6.367 | -5.9 | 69.8 | -1.59 | 0.113 |
| PhD num. collaborators | Unmatched | 8.65 | 8.775 | -1.8 | — | -0.56 | 0.573 |
| | Matched | 8.657 | 8.728 | -1 | 42.7 | -0.27 | 0.791 |
| Woman advisor | Unmatched | 0.199 | 0.124 | 20.4 | — | 6.72 | 0 |
| | Matched | 0.199 | 0.146 | 14.5 | 28.9 | 3.56 | 0 |
| Seniority Q2 | Unmatched | 0.224 | 0.212 | 2.9 | — | 0.92 | 0.358 |
| | Matched | 0.224 | 0.218 | 1.5 | 50.3 | 0.37 | 0.714 |
| Seniority Q3 | Unmatched | 0.228 | 0.238 | -2.3 | — | -0.72 | 0.473 |
| | Matched | 0.228 | 0.234 | -1.3 | 42.8 | -0.33 | 0.738 |
| Seniority Q4 | Unmatched | 0.231 | 0.224 | 1.7 | — | 0.52 | 0.602 |
| | Matched | 0.231 | 0.229 | 0.5 | 72.7 | 0.11 | 0.909 |
| Seniority Q5 | Unmatched | 0.173 | 0.163 | 2.8 | — | 0.89 | 0.371 |
| | Matched | 0.173 | 0.17 | 0.8 | 70.3 | 0.21 | 0.833 |
| Num. Papers of Advisors | Unmatched | 13.55 | 13.292 | 1.4 | — | 0.43 | 0.669 |
| | Matched | 13.554 | 13.394 | 0.9 | 38 | 0.22 | 0.828 |
| **With Ph.D. interdisciplinarity** | | | | | | | |
| PhD interdisciplinarity | Unmatched | 0.444 | 0.439 | 2.8 | — | 0.9 | 0.369 |
| | Matched | 0.444 | 0.441 | 1.8 | 35.8 | 0.45 | 0.65 |
| PhD univ. ranking | Unmatched | 83.455 | 83.878 | -2.5 | — | -0.79 | 0.428 |
| | Matched | 83.442 | 83.67 | -1.4 | 46.3 | -0.34 | 0.733 |
| PhD avg. citations | Unmatched | 1.677 | 1.733 | -2.4 | — | -0.77 | 0.444 |
| | Matched | 1.667 | 1.708 | -1.8 | 25.9 | -0.46 | 0.647 |
| PhD num. papers | Unmatched | 6.122 | 6.965 | -19.7 | — | -6.03 | 0 |
| | Matched | 6.112 | 6.368 | -6 | 69.7 | -1.59 | 0.112 |
| PhD num. collaborators | Unmatched | 8.65 | 8.775 | -1.8 | — | -0.56 | 0.573 |
| | Matched | 8.657 | 8.728 | -1 | 43.3 | -0.26 | 0.793 |
| Woman advisor | Unmatched | 0.199 | 0.124 | 20.4 | — | 6.72 | 0 |
| | Matched | 0.199 | 0.146 | 14.4 | 29.2 | 3.55 | 0 |
| Seniority Q2 | Unmatched | 0.224 | 0.212 | 2.9 | — | 0.92 | 0.358 |
| | Matched | 0.224 | 0.218 | 1.5 | 49.9 | 0.37 | 0.712 |
| Seniority Q3 | Unmatched | 0.228 | 0.238 | -2.3 | — | -0.72 | 0.473 |
| | Matched | 0.228 | 0.234 | -1.3 | 43 | -0.33 | 0.739 |
| Seniority Q4 | Unmatched | 0.231 | 0.224 | 1.7 | — | 0.52 | 0.602 |
| | Matched | 0.231 | 0.229 | 0.4 | 73.4 | 0.11 | 0.911 |
| Seniority Q5 | Unmatched | 0.173 | 0.163 | 2.8 | — | 0.89 | 0.371 |
| | Matched | 0.173 | 0.17 | 0.9 | 70 | 0.21 | 0.83 |
| Num. Papers of Advisors | Unmatched | 13.55 | 13.292 | 1.4 | — | 0.43 | 0.669 |
| | Matched | 13.554 | 13.398 | 0.8 | 39.2 | 0.21 | 0.832 |
| **Life & Earth** | | | | | | | |
| PhD univ. ranking | Unmatched | 78.864 | 77.432 | 7.3 | — | 2.07 | 0.038 |
| | Matched | 78.858 | 78.749 | 0.6 | 92.4 | 0.14 | 0.887 |
| PhD avg. citations | Unmatched | 1.354 | 1.46 | -6.4 | — | -1.78 | 0.075 |
| | Matched | 1.354 | 1.372 | -1.1 | 82.7 | -0.31 | 0.756 |
| PhD num. papers | Unmatched | 6.038 | 7.222 | -27.5 | — | -7.82 | 0 |
| | Matched | 6.042 | 6.241 | -4.6 | 83.2 | -1.23 | 0.22 |
| PhD num. collaborators | Unmatched | 9.17 | 10.408 | -17.4 | — | -4.94 | 0 |
| | Matched | 9.177 | 9.4 | -3.1 | 82 | -0.81 | 0.417 |
| Woman advisor | Unmatched | 0.246 | 0.148 | 24.6 | — | 7.29 | 0 |
| | Matched | 0.245 | 0.213 | 8 | 67.3 | 1.88 | 0.06 |
| Seniority Q2 | Unmatched | 0.179 | 0.173 | 1.6 | — | 0.46 | 0.645 |
| | Matched | 0.179 | 0.177 | 0.6 | 60.1 | 0.16 | 0.874 |
| Seniority Q3 | Unmatched | 0.253 | 0.247 | 1.3 | — | 0.38 | 0.702 |
| | Matched | 0.253 | 0.253 | 0 | 98 | -0.01 | 0.995 |
| Seniority Q4 | Unmatched | 0.225 | 0.234 | -2.2 | — | -0.62 | 0.535 |
| | Matched | 0.226 | 0.227 | -0.4 | 80.8 | -0.1 | 0.917 |
| Seniority Q5 | Unmatched | 0.221 | 0.223 | -0.3 | — | -0.1 | 0.923 |
| | Matched | 0.221 | 0.223 | -0.5 | -49.7 | -0.12 | 0.901 |
| Num. Papers of Advisors | Unmatched | 14.562 | 16.436 | -8.2 | — | -2.3 | 0.022 |
| | Matched | 14.569 | 15.125 | -2.4 | 70.3 | -0.63 | 0.526 |
| **With Ph.D. interdisciplinarity** | | | | | | | |
| PhD interdisciplinarity | Unmatched | 0.463 | 0.429 | 23.5 | — | 6.75 | 0 |
| | Matched | 0.463 | 0.457 | 4 | 82.9 | 1.01 | 0.312 |
| PhD univ. ranking | Unmatched | 78.864 | 77.432 | 7.3 | — | 2.07 | 0.038 |
| | Matched | 78.858 | 78.775 | 0.4 | 94.2 | 0.11 | 0.914 |
| PhD avg. citations | Unmatched | 1.354 | 1.46 | -6.4 | — | -1.78 | 0.075 |
| | Matched | 1.354 | 1.369 | -0.9 | 85.6 | -0.27 | 0.79 |
| PhD num. papers | Unmatched | 6.038 | 7.222 | -27.5 | — | -7.82 | 0 |
| | Matched | 6.042 | 6.222 | -4.2 | 84.8 | -1.11 | 0.268 |
| PhD num. collaborators | Unmatched | 9.17 | 10.408 | -17.4 | — | -4.94 | 0 |
| | Matched | 9.177 | 9.394 | -3 | 82.5 | -0.79 | 0.431 |
| Woman advisor | Unmatched | 0.246 | 0.148 | 24.6 | — | 7.29 | 0 |
| | Matched | 0.245 | 0.215 | 7.7 | 68.6 | 1.8 | 0.072 |
| Seniority Q2 | Unmatched | 0.179 | 0.173 | 1.6 | — | 0.46 | 0.645 |
| | Matched | 0.179 | 0.178 | 0.4 | 74.8 | 0.1 | 0.921 |
| Seniority Q3 | Unmatched | 0.253 | 0.247 | 1.3 | — | 0.38 | 0.702 |
| | Matched | 0.253 | 0.252 | 0.2 | 85 | 0.05 | 0.961 |
| Seniority Q4 | Unmatched | 0.225 | 0.234 | -2.2 | — | -0.62 | 0.535 |
| | Matched | 0.226 | 0.228 | -0.5 | 76.3 | -0.13 | 0.898 |
| Seniority Q5 | Unmatched | 0.221 | 0.223 | -0.3 | — | -0.1 | 0.923 |
| | Matched | 0.221 | 0.223 | -0.5 | -50.7 | -0.13 | 0.9 |
| Num. Papers of Advisors | Unmatched | 14.562 | 16.436 | -8.2 | — | -2.3 | 0.022 |
| | Matched | 14.569 | 15.192 | -2.7 | 66.8 | -0.7 | 0.485 |
| **Biology & Health** | | | | | | | |
| PhD univ. ranking | Unmatched | 77.435 | 77.436 | 0 | — | 0 | 0.999 |
| | Matched | 77.435 | 77.283 | 0.8 | -24017.6 | 0.33 | 0.74 |
| PhD avg. citations | Unmatched | 1.074 | 1.245 | -13.4 | — | -5.46 | 0 |
| | Matched | 1.074 | 1.093 | -1.5 | 88.7 | -0.74 | 0.457 |
| PhD num. papers | Unmatched | 5.924 | 6.909 | -22.9 | — | -9.36 | 0 |
| | Matched | 5.924 | 6.083 | -3.7 | 83.9 | -1.6 | 0.111 |
| PhD num. collaborators | Unmatched | 9.807 | 10.766 | -12.8 | — | -5.21 | 0 |
| | Matched | 9.807 | 10.06 | -3.4 | 73.6 | -1.42 | 0.155 |
| Woman advisor | Unmatched | 0.454 | 0.226 | 49.4 | — | 20.12 | 0 |
| | Matched | 0.454 | 0.434 | 4.3 | 91.2 | 1.65 | 0.098 |
| Seniority Q2 | Unmatched | 0.181 | 0.188 | -1.8 | — | -0.72 | 0.473 |
| | Matched | 0.181 | 0.181 | 0 | 99 | -0.01 | 0.994 |
| Seniority Q3 | Unmatched | 0.232 | 0.236 | -1.1 | — | -0.46 | 0.648 |
| | Matched | 0.232 | 0.228 | 0.8 | 29.2 | 0.33 | 0.742 |
| Seniority Q4 | Unmatched | 0.23 | 0.249 | -4.5 | — | -1.82 | 0.068 |
| | Matched | 0.23 | 0.242 | -2.9 | 35.2 | -1.2 | 0.231 |
| Seniority Q5 | Unmatched | 0.172 | 0.18 | -2.2 | — | -0.89 | 0.374 |
| | Matched | 0.172 | 0.177 | -1.3 | 39.8 | -0.54 | 0.588 |
| Num. Papers of Advisors | Unmatched | 10.611 | 11.275 | -4.4 | — | -1.78 | 0.075 |
| | Matched | 10.611 | 11.051 | -2.9 | 33.7 | -1.14 | 0.253 |
| **With Ph.D. interdisciplinarity** | | | | | | | |
| PhD interdisciplinarity | Unmatched | 0.46 | 0.423 | 30.8 | — | 12.55 | 0 |
| | Matched | 0.46 | 0.454 | 5.5 | 82.2 | 2.3 | 0.021 |
| PhD univ. ranking | Unmatched | 77.435 | 77.436 | 0 | — | 0 | 0.999 |
| | Matched | 77.435 | 77.196 | 1.3 | -37850.6 | 0.52 | 0.602 |
| PhD avg. citations | Unmatched | 1.074 | 1.245 | -13.4 | — | -5.46 | 0 |
| | Matched | 1.074 | 1.081 | -0.6 | 95.5 | -0.3 | 0.766 |
| PhD num. papers | Unmatched | 5.924 | 6.909 | -22.9 | — | -9.36 | 0 |
| | Matched | 5.924 | 6.068 | -3.3 | 85.5 | -1.43 | 0.153 |
| PhD num. collaborators | Unmatched | 9.807 | 10.766 | -12.8 | — | -5.21 | 0 |
| | Matched | 9.807 | 10.028 | -2.9 | 77 | -1.24 | 0.217 |
| Woman advisor | Unmatched | 0.454 | 0.226 | 49.4 | — | 20.12 | 0 |
| | Matched | 0.454 | 0.429 | 5.4 | 89 | 2.07 | 0.039 |
| Seniority Q2 | Unmatched | 0.181 | 0.188 | -1.8 | — | -0.72 | 0.473 |
| | Matched | 0.181 | 0.18 | 0.4 | 78.7 | 0.16 | 0.876 |
| Seniority Q3 | Unmatched | 0.232 | 0.236 | -1.1 | — | -0.46 | 0.648 |
| | Matched | 0.232 | 0.228 | 0.8 | 28.4 | 0.33 | 0.74 |
| Seniority Q4 | Unmatched | 0.23 | 0.249 | -4.5 | — | -1.82 | 0.068 |
| | Matched | 0.23 | 0.242 | -2.9 | 34.7 | -1.21 | 0.227 |
| Seniority Q5 | Unmatched | 0.172 | 0.18 | -2.2 | — | -0.89 | 0.374 |
| | Matched | 0.172 | 0.175 | -1 | 55.1 | -0.41 | 0.685 |
| Num. Papers of Advisors | Unmatched | 10.611 | 11.275 | -4.4 | — | -1.78 | 0.075 |
| | Matched | 10.611 | 10.879 | -1.8 | 59.5 | -0.71 | 0.477 |
| **Social Sciences** | | | | | | | |
| PhD univ. ranking | Unmatched | 81.352 | 82.048 | -4.2 | — | -1.89 | 0.058 |
| | Matched | 81.349 | 81.717 | -2.2 | 47.1 | -0.98 | 0.327 |
| PhD avg. citations | Unmatched | 1.461 | 1.634 | -6.9 | — | -3.05 | 0.002 |
| | Matched | 1.447 | 1.505 | -2.3 | 66.3 | -1.62 | 0.105 |
| PhD num. papers | Unmatched | 4.407 | 4.401 | 0.1 | — | 0.07 | 0.947 |
| | Matched | 4.408 | 4.329 | 2.1 | -1325.5 | 0.94 | 0.347 |
| PhD num. collaborators | Unmatched | 5.953 | 5.427 | 9.4 | — | 4.22 | 0 |
| | Matched | 5.953 | 5.637 | 5.7 | 39.9 | 2.46 | 0.014 |
| Woman advisor | Unmatched | 0.426 | 0.241 | 40 | — | 17.98 | 0 |
| | Matched | 0.426 | 0.41 | 3.4 | 91.6 | 1.4 | 0.163 |
| Seniority Q2 | Unmatched | 0.207 | 0.189 | 4.4 | — | 1.99 | 0.047 |
| | Matched | 0.207 | 0.198 | 2.4 | 46.1 | 1.05 | 0.296 |
| Seniority Q3 | Unmatched | 0.198 | 0.211 | -3.1 | — | -1.41 | 0.158 |
| | Matched | 0.198 | 0.21 | -3 | 5.9 | -1.31 | 0.191 |
| Seniority Q4 | Unmatched | 0.187 | 0.192 | -1.2 | — | -0.52 | 0.604 |
| | Matched | 0.187 | 0.184 | 0.9 | 22.1 | 0.4 | 0.689 |
| Seniority Q5 | Unmatched | 0.181 | 0.174 | 1.9 | — | 0.84 | 0.404 |
| | Matched | 0.181 | 0.171 | 2.7 | -43.2 | 1.18 | 0.238 |
| Num. Papers of Advisors | Unmatched | 5.84 | 5.717 | 1.4 | — | 0.61 | 0.54 |
| | Matched | 5.795 | 5.559 | 2.6 | -92.5 | 1.18 | 0.238 |
| **With Ph.D. interdisciplinarity** | | | | | | | |
| PhD interdisciplinarity | Unmatched | 0.469 | 0.421 | 32.1 | — | 14.36 | 0 |
| | Matched | 0.469 | 0.462 | 4.3 | 86.5 | 2.05 | 0.041 |
| PhD univ. ranking | Unmatched | 81.352 | 82.048 | -4.2 | — | -1.89 | 0.058 |
| | Matched | 81.352 | 81.412 | -0.4 | 91.4 | -0.16 | 0.875 |
| PhD avg. citations | Unmatched | 1.461 | 1.634 | -6.9 | — | -3.05 | 0.002 |
| | Matched | 1.461 | 1.491 | -1.2 | 82.5 | -0.78 | 0.437 |
| PhD num. papers | Unmatched | 4.407 | 4.401 | 0.1 | — | 0.07 | 0.947 |
| | Matched | 4.407 | 4.441 | -0.9 | -508.8 | -0.4 | 0.689 |
| PhD num. collaborators | Unmatched | 5.953 | 5.427 | 9.4 | — | 4.22 | 0 |
| | Matched | 5.953 | 5.87 | 1.5 | 84.4 | 0.63 | 0.527 |
| Woman advisor | Unmatched | 0.426 | 0.241 | 40 | — | 17.98 | 0 |
| | Matched | 0.426 | 0.396 | 6.5 | 83.6 | 2.71 | 0.007 |
| Seniority Q2 | Unmatched | 0.207 | 0.189 | 4.4 | — | 1.99 | 0.047 |
| | Matched | 0.207 | 0.202 | 1.2 | 73.5 | 0.51 | 0.608 |
| Seniority Q3 | Unmatched | 0.198 | 0.211 | -3.1 | — | -1.41 | 0.158 |
| | Matched | 0.198 | 0.203 | -1.1 | 64.1 | -0.5 | 0.615 |
| Seniority Q4 | Unmatched | 0.187 | 0.192 | -1.2 | — | -0.52 | 0.604 |
| | Matched | 0.187 | 0.186 | 0.2 | 82.7 | 0.09 | 0.929 |
| Seniority Q5 | Unmatched | 0.181 | 0.174 | 1.9 | — | 0.84 | 0.404 |
| | Matched | 0.181 | 0.181 | 0.1 | 93.5 | 0.05 | 0.957 |
| Num. Papers of Advisors | Unmatched | 5.84 | 5.717 | 1.4 | — | 0.61 | 0.54 |
| | Matched | 5.84 | 5.833 | 0.1 | 94.8 | 0.03 | 0.976 |

**Table S6. Field and subfield classification and prediction performance.**

| Field | Subfield | Precision | Recall | F1 score | Frequency in the test set | % |
|---|---|---|---|---|---|---|
| Math & Computing | Computing | 0.93 | 0.87 | 0.9 | 141 | 4.7 |
| | Math | 0.89 | 0.92 | 0.91 | 131 | 4.4 |
| Phys & Eng | Architecture | 0.65 | 0.71 | 0.68 | 24 | 0.8 |
| | Engineering | 0.95 | 0.92 | 0.94 | 379 | 12.7 |
| | Physics | 0.84 | 0.9 | 0.87 | 69 | 2.3 |
| Life & Earth | Agriculture | 0.9 | 0.78 | 0.83 | 45 | 1.5 |
| | Chemistry | 0.92 | 0.94 | 0.93 | 84 | 2.8 |
| | Earth | 0.81 | 0.79 | 0.8 | 178 | 6 |
| Bio & Health | Biology | 0.73 | 0.83 | 0.78 | 149 | 5 |
| | Health | 0.87 | 0.84 | 0.86 | 268 | 9 |
| | Medicine | 0.83 | 0.79 | 0.81 | 103 | 3.4 |
| Social Sciences | Anthropology | 0.83 | 0.84 | 0.84 | 45 | 1.5 |
| | Business | 0.82 | 0.86 | 0.84 | 83 | 2.8 |
| | Economics | 0.85 | 0.92 | 0.88 | 61 | 2 |
| | Education | 0.83 | 0.79 | 0.81 | 256 | 8.6 |
| | Journalism | 0.88 | 0.89 | 0.88 | 114 | 3.8 |
| | Politics | 0.87 | 0.91 | 0.89 | 87 | 2.9 |
| | Psychology | 0.8 | 0.88 | 0.84 | 228 | 7.6 |
| | Sociology | 0.84 | 0.76 | 0.8 | 165 | 5.5 |
| Humanities | Arts | 0.86 | 0.88 | 0.87 | 64 | 2.1 |
| | History | 0.91 | 0.84 | 0.87 | 74 | 2.5 |
| | Literature | 0.86 | 0.86 | 0.86 | 171 | 5.7 |
| | Philosophy | 0.93 | 0.96 | 0.94 | 26 | 0.9 |
| | Religion | 0.82 | 0.93 | 0.87 | 45 | 1.5 |

**Table S7. Odds ratios and standard errors of Ph.D. interdisciplinarity in logistic regression of placement types for faculty placement at top universities by fields, controlling various collaboration network measures.** Model specifications follow **Fig. 1.** *** $p<0.001$; ** $p<0.01$; * $p<0.05$.

| | **Math & Computing** | **Phys & Eng** | **Life & Earth** | **Bio & Health** | **Social Sciences** |
|---|---|---|---|---|---|
| Full controls | 0.985 (0.077) | 0.925 (0.043) | 0.840* (0.058) | 0.837*** (0.037) | 0.882* (0.046) |
| +Betweenness centrality | 0.971 (0.077) | 0.926 (0.043) | 0.839* (0.058) | 0.837*** (0.037) | 0.886* (0.046) |
| +Closeness centrality | 0.983 (0.077) | 0.925 (0.043) | 0.838* (0.058) | 0.836*** (0.037) | 0.883* (0.046) |
| +Degree centrality | 0.989 (0.078) | 0.922 (0.043) | 0.837** (0.058) | 0.837*** (0.037) | 0.881* (0.046) |
| +Eigenvector centrality | 0.985 (0.080) | 0.919* (0.043) | 0.834** (0.058) | 0.835*** (0.037) | 0.885* (0.046) |